\documentclass[iop, numberedappendix]{emulateapj}
\usepackage{amssymb}
\usepackage{amsmath}
\usepackage{float}
\usepackage{relsize}
\usepackage{verbatim}
\usepackage{color}
\usepackage{comment}
\usepackage{graphicx}
\usepackage{bm}

\newcommand{\todo}[1]{}

\def\be{\begin{equation}}
\def\ee{\end{equation}}
\def\ba{\begin{eqnarray}}
\def\ea{\end{eqnarray}}

\newcommand{\DM}{\ensuremath{\mathrm{DM}}} %\xspace
\newcommand{\RP}{\ensuremath{\mathrm{RP}}} %\xspace
\newcommand{\FD}{\ensuremath{\Delta t_{\mathrm{FD},j}}}
\newcommand{\SN}{\ensuremath{\mathrm{S/N}}}

\newcommand{\J}{\ensuremath{\mathrm{J}}}

\newcommand{\const}{\ensuremath{\mathrm{const}}}
\newcommand{\red}{\ensuremath{\mathrm{r}}} 
\newcommand{\DISS}{\ensuremath{\mathrm{DISS}}}

\newcommand{\Like}{\ensuremath{\mathcal{L}}}
\newcommand{\R}{\ensuremath{\mathcal{R}}}

\newcommand{\UL}{\ensuremath{\mathrm{UL}}}
\newcommand{\N}{\ensuremath{\mathcal{N}}}

\newcommand{\cmthree}{\ensuremath{\mathrm{cm^{-3}}}}
\newcommand{\widebar}{\overline}
\newcommand{\s}{s}

\newcommand{\niss}{n_{\rm ISS}}

\newcommand{\Dtd}{\Delta t_{\mathrm{d}}}
\newcommand{\Dnud}{\Delta \nu_{\mathrm{d}}}

\newcommand{\taud}{\tau_{\mathrm{d}}}

\newcommand{\tempo}{{\textsc{tempo}}}
\newcommand{\tempotwo}{{\textsc{tempo2}}}
\newcommand{\tobs}{\ensuremath{T_{\mathrm{obs}}}}
\newcommand{\Bchan}{\ensuremath{B_{\mathrm{chan}}}}
\newcommand{\nuband}{\ensuremath{\nu_{\mathrm{band}}}}
\newcommand{\ex}{{\rm ex}}
\newcommand{\sigex}{{\sigma_\ex}}

\newcommand{\Ared}{{A_{\red}}}
\newcommand{\gammared}{{\gamma_{\red}}}
\newcommand{\alphared}{{\alpha_{\red}}}

\shorttitle{Excess Noise in NANOGrav Pulsars}
\shortauthors{Lam et al.}

\begin{document}

\title{
The NANOGrav Nine-Year Data Set: Excess Noise in Millisecond Pulsar Arrival Times
}
\author{ 
M.\,T.\,Lam\altaffilmark{1,2},
J.\,M.\,Cordes\altaffilmark{2},
S.\,Chatterjee\altaffilmark{2}, 
Z.\,Arzoumanian\altaffilmark{3},
K.\,Crowter\altaffilmark{4},
P.\,B.\,Demorest\altaffilmark{5},
T.\,Dolch\altaffilmark{6},
J.\,A.\,Ellis\altaffilmark{7,20},
R.\,D.\,Ferdman\altaffilmark{8,9},
E.\,Fonseca\altaffilmark{4},
M.\,E.\,Gonzalez\altaffilmark{4,10},
G.\,Jones\altaffilmark{11},
M.\,L.\,Jones\altaffilmark{1},
L.\,Levin\altaffilmark{12},
D.\,R.\,Madison\altaffilmark{13},
M.\,A.\,McLaughlin\altaffilmark{1},
D.\,J.\,Nice\altaffilmark{14},
T.\,T.\,Pennucci\altaffilmark{1,11},
S.\,M.\,Ransom\altaffilmark{13},
R.\,M.\,Shannon\altaffilmark{15,16}, 
X.\,Siemens\altaffilmark{17},
I.\,H.\,Stairs\altaffilmark{4},
K.\,Stovall\altaffilmark{5,18},
J.\,K.\,Swiggum\altaffilmark{17},
W.\,W.\,Zhu\altaffilmark{19}}
\altaffiltext{1}{Department of Physics and Center for Gravitational Waves and Cosmology, West Virginia University, White Hall, Morgantown, WV 26506, USA; michael.lam@mail.wvu.edu}
\altaffiltext{2}{Department of Astronomy and Cornell Center for Astrophysics and Planetary Science, Cornell University, Ithaca, NY 14853, USA}
\altaffiltext{3}{Center for Research and Exploration in Space Science and Technology and X-Ray Astrophysics Laboratory, NASA Goddard Space Flight Center, Code 662, Greenbelt, MD 20771, USA}
\altaffiltext{4}{Department of Physics and Astronomy, University of British Columbia, 6224 Agricultural Road, Vancouver, BC V6T 1Z1, Canada}
\altaffiltext{5}{National Radio Astronomy Observatory, P.O.~Box 0, Socorro, NM, 87801, USA}
\altaffiltext{6}{Department of Physics, Hillsdale College, 33 E. College Street, Hillsdale, MI 49242, USA}
\altaffiltext{7}{Jet Propulsion Laboratory, California Institute of Technology, 4800 Oak Grove Dr. Pasadena CA, 91109, USA}
\altaffiltext{8}{Department of Physics, McGill University, 3600 rue Universite, Montreal, QC H3A 2T8, Canada}
\altaffiltext{9}{Department of Physics, University of East Anglia, Norwich NR4 7TJ, United Kingdom}
\altaffiltext{10}{Department of Nuclear Medicine, Vancouver Coastal Health Authority, Vancouver, BC V5Z 1M9, Canada}
\altaffiltext{11}{Department of Physics, Columbia University, 550 W. 120th St. New York, NY 10027, USA}
\altaffiltext{12}{Jodrell Bank Centre for Astrophysics, School of Physics and Astronomy, The University of Manchester, Manchester M13 9PL, UK}
\altaffiltext{13}{National Radio Astronomy Observatory, 520 Edgemont Road, Charlottesville, VA 22903, USA}

\begin{abstract}

%217 words
Gravitational wave astronomy using a pulsar timing array requires high-quality millisecond pulsars, correctable interstellar propagation delays, and high-precision measurements of pulse times of arrival.  Here we identify noise in timing residuals that exceeds that predicted for arrival time estimation for millisecond pulsars observed by the North American Nanohertz Observatory for Gravitational Waves. We characterize the excess noise using variance and structure function analyses. We find that 26 out of 37 pulsars show inconsistencies with a white-noise-only model based on the short timescale analysis of each pulsar and we demonstrate that the excess noise has a red power spectrum for 15 pulsars. We also decompose the excess noise into chromatic (radio-frequency-dependent) and achromatic components. Associating the achromatic red-noise component with spin noise and including additional power-spectrum-based estimates from the literature, we estimate a scaling law in terms of spin parameters (frequency and frequency derivative) and data-span length and compare it to the scaling law of Shannon \& Cordes (2010).  We briefly discuss our results in terms of detection of gravitational waves at nanohertz frequencies.

\end{abstract}


\section{Introduction}

Recycled millisecond pulsars (MSPs) are the most stable astrophysical clocks in the Universe \citep{Verbiest+2009}. Such stability allows them to be used as tools in experiments of fundamental physics, including tests of gravitation \citep[e.g.,][]{Zhu+2015}, constraints on general relativity (see \citealt{Will2014} for an overview), and detection of nHz-$\mu$Hz-frequency gravitational waves (GWs; e.g., \citealt{NG5BWM,NG9GWB,Verbiest+2016}).

\footnotetext[14]{Department of Physics, Lafayette College, Easton, PA 18042, USA}
\footnotetext[15]{CSIRO Astronomy and Space Science, Australia Telescope National Facility, Box 76, Epping NSW 1710, Australia}
\footnotetext[16]{International Centre for Radio Astronomy Research, Curtin University, Bentley WA 6102, Australia}
\footnotetext[17]{Center for Gravitation, Cosmology and Astrophysics, Department of Physics, University of Wisconsin-Milwaukee, P.O. Box 413, Milwaukee, WI 53201, USA}
\footnotetext[18]{Department of Physics and Astronomy, University of New Mexico, Albuquerque, NM, 87131, USA}
\footnotetext[19]{Max-Planck-Institut f\"{u}r Radioastronomie, Auf dem H\"{u}gel 69, D-53121, Bonn, Germany}
\footnotetext[20]{Einstein Fellow}

GWs will appear as temporally-correlated stochastic deviations from a timing model that includes kinematic and astrometric terms that describe the pulsar-observatory line of sight \citep{Sazhin1978,Detweiler1979,hd1983,Romani1989,fb1990}, as well as frequency-dependent pulse profile evolution and interstellar propagation effects \citep{Craft1970,Armstrong1984, brn1984, Rickett1990, fc1990, cs2010}. Detection of GWs is accomplished by measuring a specific, quadrupolar angular correlation in pulse time-of-arrival (TOA) deviations from many pulsars distributed over the sky \citep{hd1983}. However, numerous other sources of noise, both temporally uncorrelated and correlated, contaminate our timing models and reduce each pulsar's contribution to the detection of GWs \citep[e.g.,][]{NG9yr,Lentati+2016}. Fortunately these other sources are uncorrelated between pulsars.

\setcounter{footnote}{0}

\citet[][hereafter SC2010]{sc2010} investigated the strength and non-stationarity of long-term, correlated ``red'' noise in TOAs from pulsars in the literature. They attributed the red noise to rotational instabilities in individual pulsars, known as spin noise. Arguing that spin noise was the dominant source of red noise in the pulsars, they globally characterized and developed scaling relations of spin noise over all pulsar populations, finding that the spin noise was a steep-spectrum process ($\sim f^{-5}$) in a fit dominated by slow-period pulsars. 

Other sources of red noise in TOAs are known. One such source is the interstellar medium (ISM), which causes radio-frequency-dependent delays in pulse arrival times \citep{Lam+2016}. The largest effect comes from a dispersive delay proportional to the dispersion measure (DM), the electron density integral along the line of sight, and inverse radio frequency squared. The DM is typically estimated using multi-frequency observations and DM delays are removed by imposing some correlation structure in time \citep{Keith+2013,NG9yr}. Imperfect estimation of DM, from using incorrect temporal correlations, from combination of asynchronously observed multi-frequency data, or from frequency-dependent DMs due to interstellar scattering, will cause red noise in the timing residuals\footnote{defined as TOAs minus timing model} \citep{Lam+2015,css2016,Lam+2016}. Other noise sources that can appear temporally correlated are possible, including variations in the pulse shape from interstellar scattering, polarization calibration (e.g., cross-coupling) errors, radio frequency interference (RFI), and pulsar mode changes \citep[see Appendix A of][hereafter NG9WN]{NG9WN}.

The North American Nanohertz Observatory for Gravitational Waves \citep[NANOGrav;][]{McLaughlin2013} collaboration observes a pulsar timing array for the detection and long-term study of GWs. The goal is to correlate arrival time differences in pulses from individual pulsars after a timing model and noise model have been accounted for each pulsar. Understanding the noise model is crucial for properly calibrating the array for optimal GW sensitivity \citep{NG9WN}. This work aims to characterize the timing accuracy of the NANOGrav MSPs, paralleling the results of NG9WN to characterize their timing precision. Throughout this work, we will denote long-term ($\sim$ years) temporally-correlated noise as ``red noise'' in reference to spectral power concentrated at low-frequencies. We will denote changes of noise properties with radio observing frequency as ``chromatic''.

In \S\ref{sec:excess_variance_methodology}, we describe the methodology for estimating the variance from stochastic noise processes. In \S\ref{sec:observations}, we describe the NANOGrav nine-year data set. We describe the implementation of the measured white-noise model for producing residuals and estimate the excess noise in our MSPs in \S\ref{sec:excess_variance_estimation}; we discuss the pulsars individually in \S\ref{sec:discussion}. In \S\ref{sec:MLA}, we insert our measurements into the SC2010 data set and re-derive the scaling relations for red noise in various pulsar class subsets. We summarize our conclusions in \S\ref{sec:conclusions}. For convenience, we provide a table of symbols used in the paper in Table~\ref{table:definitions}.

\begin{deluxetable*}{llc}
%\tablewidth{0.8\textwidth}
\centering
\tablecolumns{3}
\tablecaption{Symbols and Acronyms Used}
\tablehead{
\colhead{Symbols} & \colhead{Definition} & \colhead{Typical Units}\\
}
\startdata
$A$ & Achromatic excess noise (AEN)& \\ 
$\Ared$ & Coefficient in power-law amplitude spectrum & $\mathrm{\mu s~yr^{1/2}}$\\ %Amplitude of power-law power spectrum
$\Bchan$ & Channel bandwidth & MHz \\
$C$ & Chromatic excess noise (CEN) & \\ 
$\mathbf{C}$ & Noise covariance matrix & $\mathrm{\mu s}^2$\\
$C_2$ & Spin noise fitting parameter for noise amplitude & \\
DM & Dispersion Measure & pc \cmthree\\
$D_{\widebar{\R}}(\tau)$ & Structure function of epoch-averaged residuals & $\mathrm{\mu s}^2$\\
$\mathcal{F}$ & Scaling factor on error (EFAC) & \\
$f$ & Fluctuation frequency & yr$^{-1}$\\
%$f_c$ & Corner frequency for \citet{Reardon+2016} & yr$^{-1}$\\
$f_{\rm D}$ & Probability density function for detections & \\
%$f_h$ & High-frequency cutoff for power law & yr$^{-1}$\\
$f_{\rm UL}$ & Probability density function for upper limits & \\
$\FD$ & Frequency-dependent term in timing model & $\mathrm{\mu s}$\\
$i,j,k$ & Measurement indices. & \\
& For the variance analysis, $i$ and $j$ denote indices over time & \\
&  and frequency respectively. & \\
& For the spin noise analysis, $i$ labels individual measurements, & \\
& $j$ labels detections of red noise, and $k$ labels upper limits of red noise & \\
%$j$ & index over frequency & \\
$\mathcal{J}$ & Error correlated in observing frequency (ECORR) & $\mathrm{\mu s}$\\
$\Like$ & Likelihood or Pseudo-likelihood function & \\
%$m_3$ & Slope of third-order structure function from \cite{cd1985} & \\
$\niss$ & Number of scintles & \\
$N$ & Number of measurements & \\
$\N$ & Normal distribution & \\
$P(f)$ & Power spectrum & $\mathrm{\mu s^2~yr}$ or yr$^3$\\
$P_0$ & Amplitude of \citet{Reardon+2016} power spectrum & yr$^{-3}$\\
$\mathcal{Q}$ & Error added in quadrature (EQUAD) & $\mathrm{\mu s}$\\
$\RP$ & Redness parameter, $D_{\widebar{\R}}(\tau\geq 365~\mathrm{days})/D_{\widebar{\R}}(0 < \tau \leq 30~\mathrm{days})$ & \\
$\R$ & Timing residuals & $\mathrm{\mu s}$\\
$\widebar{\R}$ & Epoch-averaged residuals & $\mathrm{\mu s}$\\
$s$ & Power-law index for chromatic noise& \\
$t$ & Observing time & Time\\
$\tobs$ & Total timespan of single observation & min\\
$T$ & Total timespan of all observations & yr\\
$\mathcal{T}(f)$ & Transmission function applied to power spectra & \\
& to remove effects of timing model, normalized to unit height & \\
$V$ & Variance of a process or time series & $\mathrm{\mu s}^2$\\
$W$ & White noise (WN) & \\
$X$ & Generic process or time series & \\
$\alpha$ & Spin noise fitting parameter on spin frequency & \\
$\alphared$ & Index of modified power-law power spectrum & \\
& in \citet{Reardon+2016} & \\
$\beta$ & Spin noise fitting parameter on spin frequency derivative & \\
$\gamma$ & Spin noise fitting parameter on total timespan observed & \\
$\gammared$ & Index of power-law power spectrum, $P(f) \propto f^{-\gammared}$ & \\
$\delta$ & Spin noise fitting parameter for noise scatter & \\
$\delta_{X X'}$ & Kronecker delta function  & \\
$\Dtd$ & Diffractive scintillation timescale & s \\
$\Dnud$ & Diffractive scintillation bandwidth & MHz \\
$\eta$ & Dynamic spectrum filling factor & \\
$\nu$ & Electromagnetic (radio) frequency & MHz\\
$\nu_0$ & Reference observing frequency & MHz\\
$\nuband$ & Receiver band frequency & MHz\\
$\nu_{\rm s}$ & Pulsar spin frequency & Hz\\
$\dot{\nu}_{\rm s}$ & Pulsar spin frequency derivative & Hz/s\\
%$\rho_{\DISS}$ & Diffraction Interstellar Scintillation correlation matrix & \\
$\sigma$ & rms & $\mathrm{\mu s}$\\
$\sigma_\const$ & rms noise constant with pulse S/N,$\sigma_\const = \sqrt{\sigma_J^2 + \sigma_{\DISS}^2}$ & $\mathrm{\mu s}$\\
$\sigma_{\DISS}$ & Diffractive Interstellar Scintillation error & $\mathrm{\mu s}$\\
$\sigma_{\J}$ & Jitter error & $\mathrm{\mu s}$\\
$\sigma_{\SN}$ & Template-fitting error & $\mathrm{\mu s}$\\
$\sigma_{\red}$ & Red noise error & $\mathrm{\mu s}$\\
$\sigma_{\rm TN,2}$ & ``Timing noise'' after a quadratic fit, & $\mathrm{\mu s}$\\
& red noise measured in previous works & \\
$\sigex$ & Excess noise & $\mathrm{\mu s}$\\
$\tau$ & Time lag & day\\
$\taud$ & Scattering timescale & $\mathrm{\mu s}$
\enddata
\label{table:definitions}
\end{deluxetable*}

\section{Variance of Stochastic Processes with Additive Noise}
\label{sec:excess_variance_methodology}

Our goal is to use pulsar timing residuals to determine the amount of excess variance over the white noise (WN) in our pulse arrival times. More generally, we want to measure the posterior probability density function (PDF) of the variance of stochastic processes when combined with some other independent processes, e.g., WN. In this section, we develop the formalism for measuring the variance of multiple processes in our timing residuals. For clarity, we will define $V_X \equiv \sigma_X^2$ to be the variance of a process $X_{ij}$ with rms $\sigma_X$, where $i$ labels the measurement epoch and $j$ labels the observed radio frequency. Through this section, our formalism will apply to generic processes with or without temporal correlation structure, i.e., we do not distinguish between excess red or excess white noise.

\subsection{Single Stochastic Process}

Let the residuals $\R_{ij} = X_{ij} + W_{ij}$ be the measured process that is the sum of two Gaussian processes, where $X_{ij}$ is the stochastic noise process of interest with variance we wish to characterize and $W_{ij}$ is additive WN. We assume that both noise processes are zero mean; it therefore follows that $\R_{ij}$ will be zero mean.

Since $\R_{ij}$ is a zero mean Gaussian process then the PDF of $\R_{ij}$ is simply a normal distribution $\N(0,V_{\R_{ij}})$, where $V_{\R_{ij}}$ represents the variance of $\R_{ij}$. We can separate each $V_{\R_{ij}}$ into the sum of the two variances $V_{X_{ij}} + V_{W_{ij}}$ since $X_{ij}$ and $W_{ij}$ are independent processes and the covariance between the two is zero. We define the total excess variance $V_{\ex}$ over the WN to be
\be
V_{\ex} \equiv V_X = \left<X_{ij}^2\right> =  \frac{1}{N}\sum_{i,j} X_{ij}^2,
\ee
where $N$ is the number of measurements and again $X_{ij}$ is zero mean. The average denoted by angle brackets is taken over all indices $i$ and $j$. Thus, for each measurement $\R_{ij}$ with WN measurement error $\sigma_{W_{ij}}$, we approximate the variance as $V_{\R_{ij}} = V_{\ex} + V_{W_{ij}}$. We define the pseudo-likelihood function for $V_{\ex}$ as %(also the average variance, related trivially to the total variance of $x$)
\ba
&&\Like(V_{\ex} | \{\R_{ij}, V_{W_{ij}}\}) \equiv \nonumber\\
&&\prod_{i,j} \frac{1}{\sqrt{2\pi (V_{\ex} + V_{W_{ij}})}} \exp\left[-\frac{\R_{ij}^2}{2(V_{\ex} + V_{W_{ij}})}\right].
\label{eq:likelihood_ex}
\ea
The width of the pseudo-likelihood function (i.e., how well we can estimate $V_{\ex}$) will depend on the magnitude of the measurement errors $\sigma_{W_{ij}}$ and on the stationarity of $\sigex$. For example, any additional non-stationary noise processes, such as short-term (i.e., much less than the total data span) RFI, will increase not only the estimated $V_{\ex}$ but the error on $V_{\ex}$ as well.

The pseudo-likelihood function is not strictly a likelihood function unless all $\R_{ij}$ in Eq~\ref{eq:likelihood_ex} are independent, which is not known a priori. To test our pseudo-likelihood function as a variance estimator, we simulated realizations of excess noise with power-law spectra between $P(f) \propto f^0$ and $f^{-6}$ and added white noise of varying strengths. We were able to recover the variance of the injected excess noise, suggesting that the derived $V_{\ex}$ is a good estimate of the true excess noise variance. For brevity, we will drop the ``pseudo'' in future discussion.

\subsection{Achromatic and Chromatic Stochastic Processes}

We now consider residuals $\R_{ij} = A_{ij} + C_{ij} + W_{ij}$ where $A_{ij}$ and $C_{ij}$ are two Gaussian processes. Again, $W_{ij}$ is additive WN. We let $A_{ij} = A_i$ be a stochastic process that is achromatic (frequency-independent) and therefore only depends on time, such as rotational spin noise or GWs. We let $C_{ij}$ be a process that is chromatic, such as from interstellar scattering corrections.

We must impose some frequency dependence on $C_{ij}$, otherwise we can place all of the variance from $A_i$ into the time component of $C_{ij}$. Therefore, we define $C_{ij} \equiv C_i (\nu_j/\nu_0)^{-\s}$, where $C_i$ is the chromatic noise for epoch $i$, $\nu_0$ is a reference frequency, and we select and fix the power-law index $\s$ (e.g., $\s = 2$ for DM variations) in our analysis. As before, for a given measurement $\R_{ij}$ and the three independent processes, we can write the total variance as the sum of the three components, $V_{\R_{ij}} = V_{A_i} + V_{C_i}(\nu_j/\nu_0)^{-2\s} + V_{W_{ij}}$, where the factor of 2 multiplying $\s$ comes from the expected value of $C_{ij}$ squared. The (pseudo-)likelihood function can now be written as
\ba
&& \Like(V_A, V_C | \{\R_{ij}, \nu_j, V_{W_{ij}}\},\nu_0,s) = \nonumber \\
&&\prod_{i,j} \frac{1}{\sqrt{2\pi (V_A + V_C(\nu_j/\nu_0)^{-2s} + V_{W_{ij}})}} \times \nonumber \\
&&\exp\left[-\frac{\R_{ij}^2}{2(V_A + V_C(\nu_j/\nu_0)^{-2s} + V_{W_{ij}})}\right],
\label{eq:likelihood_ac}
\ea
where we have assumed that $V_A$ and $V_C$ represent the total of the achromatic and chromatic variances respectively.

We convert estimates of the chromatic variance using one value of $\s$ to a new value of $\s^\prime$ with
\be
C_{ij} = C_i \left(\frac{\nu_j}{\nu_0}\right)^{-\s} = C_i^\prime \left(\frac{\nu_j}{\nu_0}\right)^{-\s^\prime}.
\ee
The variance of the process $C_i^\prime$ will then be
\be
V_{C_i^\prime} = V_{C_i} \left(\frac{\nu_j}{\nu_0}\right)^{-2(\s-\s^\prime)}.
\ee
When rescaling reference frequencies from $\nu_0$ to $\nu_0^{\prime\prime}$, the variance of the process $C_i^{\prime\prime}$ will be $V_{C_i^{\prime\prime}} = V_{C_i} \left(\nu_0^{\prime\prime} / \nu_0\right)^{-2\s}$.

\section{Observational Data: The NANOGrav Nine-Year Data Set}
\label{sec:observations}

Here we will briefly summarize our data. We used TOAs and parameter files from the NANOGrav nine-year data set described in \citet[][hereafter NG9]{NG9yr} for our analysis. Observations of thirty-seven MSPs were carried out at the Green Bank Telescope (GBT) and Arecibo Observatory (AO). We used two generations of backends at each telescope, the GASP/GUPPI backends at GBT and the ASP/PUPPI backends at AO \citep{Demorest2007,drd+2008,fdr2010,Demorest+2013}, with more recent backends processing up to an order of magnitude larger bandwidth for improved pulse sensitivity. Observations of each pulsar at each epoch were made using two telescope receivers at widely separate frequencies.

Pulse profiles were folded and de-dispersed using an initial timing model with sufficient accuracy to keep any timing noise from profile shape changes at negligible levels (NG9WN). Polarization calibration and RFI excision algorithms were applied to the raw data profiles using the \textsc{psrchive}\footnote{\url{http://psrchive.sourceforge.net}, accessed via scripts available at \url{https://github.com/demorest/nanopipe}} software package \citep{Hotan+2004,vS+2012}. Prior to every pulsar observation, a broadband noise source was introduced into the two polarization signal paths between the receiver and the backend systems, which allowed for differential gain and phase calibration. We allowed for the possibility that the noise source power in the two hands of polarization was not constant with time and therefore observed a bright, unpolarized quasar once per month per telescope per receiver band to properly calibrate the noise source.

After calibration, known RFI signals were excised, followed by a filtering process that removed frequency channels in which the off-pulse variation in a 20-channel wide frequency window was four times greater than the median variation value. Throughout the data reduction process, profiles observed within a single epoch were averaged together in time to reduce the data volume at each stage. The final pulse profiles used to generated TOAs were fully time averaged with some frequency averaging (the amount dependent on bandwidth) to build pulse $\SN$. The final frequency resolution was of order a few MHz per channel, sufficient so that timing errors from frequency-dependent pulse profile evolution were negligible within each frequency channel (\citealt{pdr2014}; NG9WN). See NG9 for more details on the data processing.

Summing the two orthogonal polarization pulse profiles produced the intensity profiles used for arrival time estimation. TOAs were generated from the multi-frequency pulse profiles using a single smoothed template waveform per pulsar per telescope per receiver band and a Fourier-domain estimation algorithm \citep{Taylor1992}. Using the TOAs and an initial timing model as a starting point, we fit timing parameters describing the spin, astrometry, and environmental properties (e.g., binary) of each pulsar. We used the \tempo\footnote{\url{http://tempo.sourceforge.net}} software package for all timing parameter estimation. Spin-down and astrometric parameters were fit for all pulsars, and Keplerian parameters were fit for all binary pulsars. Post-Keplerian parameters were fit if they were deemed significant according to an F-test.

On a given epoch, we observed each pulsar over at least two receiver bands to estimate the dispersive delay due to the changing DM. We included one $\Delta \DM_i$ parameter (DMX) per epoch which described the DM difference for that epoch from some fixed reference DM.\footnote{Since different receiver bands are observed on different days in some cases, the DMX method assigns all TOAs within a given window of time to the same epoch and then one $\Delta \DM_i$ is fit over that epoch. In the case of NG9, the window is typically 14 days.} In addition, between zero and five, typically no more than two, frequency-dependent (FD) time delay terms ($\Delta t_{\mathrm{FD},j}$) were fit, modeling the time-independent pulse profile evolution as polynomials in the logarithm of observing frequency. Significance of the $\FD$ parameters was also determined by use of an F-test.

Lastly, in NG9 we used a parameterized noise model that included both white and red noise terms correlated differently in time and frequency (\citealt{NG5BWM}; NG9; \citealt{NG9GWB}). Both \citet{NG9WN} and our work take take a different approach to NG9 but a comparison is made in the next section. The noise covariance matrix used in NG9 is
\ba
\mathbf{C}_{\nu\nu',t t'} &=& \delta_{t t'}\left[\delta_{\nu\nu'}\left(\mathcal{F}^2 \sigma_{\SN}^2 + \mathcal{Q}^2\right) + \mathcal{J}^2\right] \nonumber \\
& & + \mathbf{C}_{\red}(\Ared,\gammared).
\ea
For TOAs with an associated error $\sigma_{\SN}$ from template-fitting, NG9 increased them with a scaling factor $\mathcal{F}$ (EFAC, $\approx 1$ for all pulsars, to within a factor of 2 for most) and added additional variances in quadrature: one that characterized the correlations between receiver bands due to processes such as pulse jitter or radio frequency interference ($\mathcal{J}$, ECORR), and one that characterized additional additive noise ($\mathcal{Q}$, EQUAD). Finally, in NG9 we modeled red noise correlated in time ($\mathbf{C}_\red$) as a Gaussian process with a power-law spectrum
\be
P(f) = \Ared^2 \left(\frac{f}{1~\mathrm{yr}^{-1}}\right)^{-\gammared}
\ee
where the spectral index $\gammared$ is positive\footnote{NG9 writes the spectrum in the form $P(f) \propto f^\gammared$ and lists all of the values of $\gammared$ as negative for red noise. In this work we will follow the common choice of $\gammared$ being positive for red noise.} for red noise and $\Ared$ is the amplitude of the red noise in units of $\mathrm{\mu s~yr^{1/2}}$. The normalization of $P(f)$ is such that the integral from $f=0$ to $f=\infty$ gives the noise variance. The noise parameters were estimated with a Bayesian inference package\footnote{\url{https://github.com/jellis18/PAL2}} and fit for in a joint likelihood with the timing model parameters. The uncertainties on the two red noise parameters are often sizable with large covariances between the values. Both parameter (par) and TOA (tim) files from the NG9 data release are available at \url{https://data.nanograv.org}.

\section{Excess Noise Estimation}
\label{sec:excess_variance_estimation}

Our goal is to estimate the amount of ``post-fit'' excess noise in our timing residuals beyond the WN model described in NG9WN. We adjusted the TOA uncertainties and timing models used in NG9 in order to generate timing residuals for each pulsar. We can characterize the excess noise using those residuals.

\subsection{The Reformulated Timing Model}
\label{sec:new_timing_model}

We removed three pulsars from our analysis where WN parameters could not be estimated for at least one pulsar/backend combination due to low pulse $\SN$. Otherwise, we added the jitter ($\sigma_\J$) and diffractive interstellar scintillation ($\sigma_{\DISS}$) errors from Table 3 of NG9WN in quadrature to the template-fitting errors after scaling them appropriately with time, such that 
\ba
\sigma_{\SN} &\rightarrow& \left[\sigma_{\SN}^2 + \sigma_\J^2(\nuband,\tobs)\right. \nonumber \\
& & + \left.\sigma_{\DISS}^2(\nuband,\Bchan,\tobs)\right]^{1/2},
\label{eq:adjustWN}
\ea
where $\nuband$ is the receiver band (not the frequency channel) of an individual observation, $\tobs$ is the total timespan of an individual observation, and $\Bchan$ is the channel bandwidth from which each TOA was generated (much less than the total of the observing band). We explicitly specify $\nuband$ because WN parameters are a function of observing band in NG9WN. The jitter noise $\sigma_\J$ is proportional to $\tobs^{-1/2}$ and is independent of observing bandwidth. The scintillation noise non-trivially scales with bandwidth and time according to the formula
\ba
\sigma_{\DISS} &\approx& \frac{\taud}{\sqrt{\niss}}\nonumber \\
&\approx& \taud \left[\left(1+\eta_t \frac{\tobs}{\Dtd}\right)\left(1+\eta_\nu \frac{\Bchan}{\Dnud}\right)\right]^{-1/2},
\ea 
where $\taud \propto \nu^{-4.4}$ is the scattering timescale, $\niss$ is the number of scintles, $\Dtd \propto \nu^{1.2}$ and $\Dnud \propto \nu^{4.4}$ are the scintillation bandwidth and timescale, respectively, and $T_{\rm obs}$ is the observation duration. Proportionalities with radio frequency are given for a Kolmogorov medium \citep{cr98,NE2001}. The filling factors $\eta_t, \eta_\nu$ are typically in the range 0.1 to 0.3 \citep{cs2010,Levin+2016}; we chose them to be 0.2 as in NG9WN. Jitter and scintillation noise were scaled in NG9WN to 30 minute observation durations and bandwidths equal to that of the full band (rather than the channel bandwidth), as provided in NG9. Each TOA in NG9 is approximately 20$-$30~minutes in duration but with channel bandwidths of order 1~MHz; we make sure to scale the errors appropriately for every individual TOA. For pulsars with upper limits on $\sigma_\J$, we set the value of $\sigma_\J$ we use equal to its 95\% upper limit, providing us with conservative lower bounds on the excess noise.

\begin{deluxetable*}{ccccccccc}
\tablewidth{0pt}
\tablecolumns{9}
\tablecaption{Components Included in Three Noise Models for Timing Fit}
\tablehead{
\colhead{Model} & \colhead{$\sigma_{\SN}$}  & \colhead{$\mathcal{F}$} & \colhead{$\mathcal{Q}$} & \colhead{$\mathcal{J}$} & \colhead{$\Ared$,$\gammared$} & \colhead{$\sigma_{\J}$}  & \colhead{$\sigma_{\DISS}$}  & \colhead{Fitting Method}
}
\startdata
A & $\times$ & $\times$ & $\times$ & $\times$ & $\times$ & & & \tempo\\
B & $\times$ & & & & & & & Iterative \tempotwo~via libstempo \\
C & $\times$ & &  & & & $\times$ & $\times$ & Iterative \tempotwo~via libstempo
\enddata
\label{table:models}
\tablenotetext{}{Columns are: model label, template-fitting error ($\sigma_{\SN}$), TOA error scaling factor (EFAC, $\mathcal{F}$), TOA additive error (EQUAD, $\mathcal{Q}$), TOA frequency-correlated error (ECORR, $\mathcal{J}$), power-law red-noise parameters ($\Ared,\gammared$), jitter error ($\sigma_{\J}$), diffractive interstellar scintillation error ($\sigma_{\DISS}$), and fitting method.}
\end{deluxetable*}

\begin{figure*}[t!]
\epsscale{1.0} %1.2
\begin{center}
\plotone{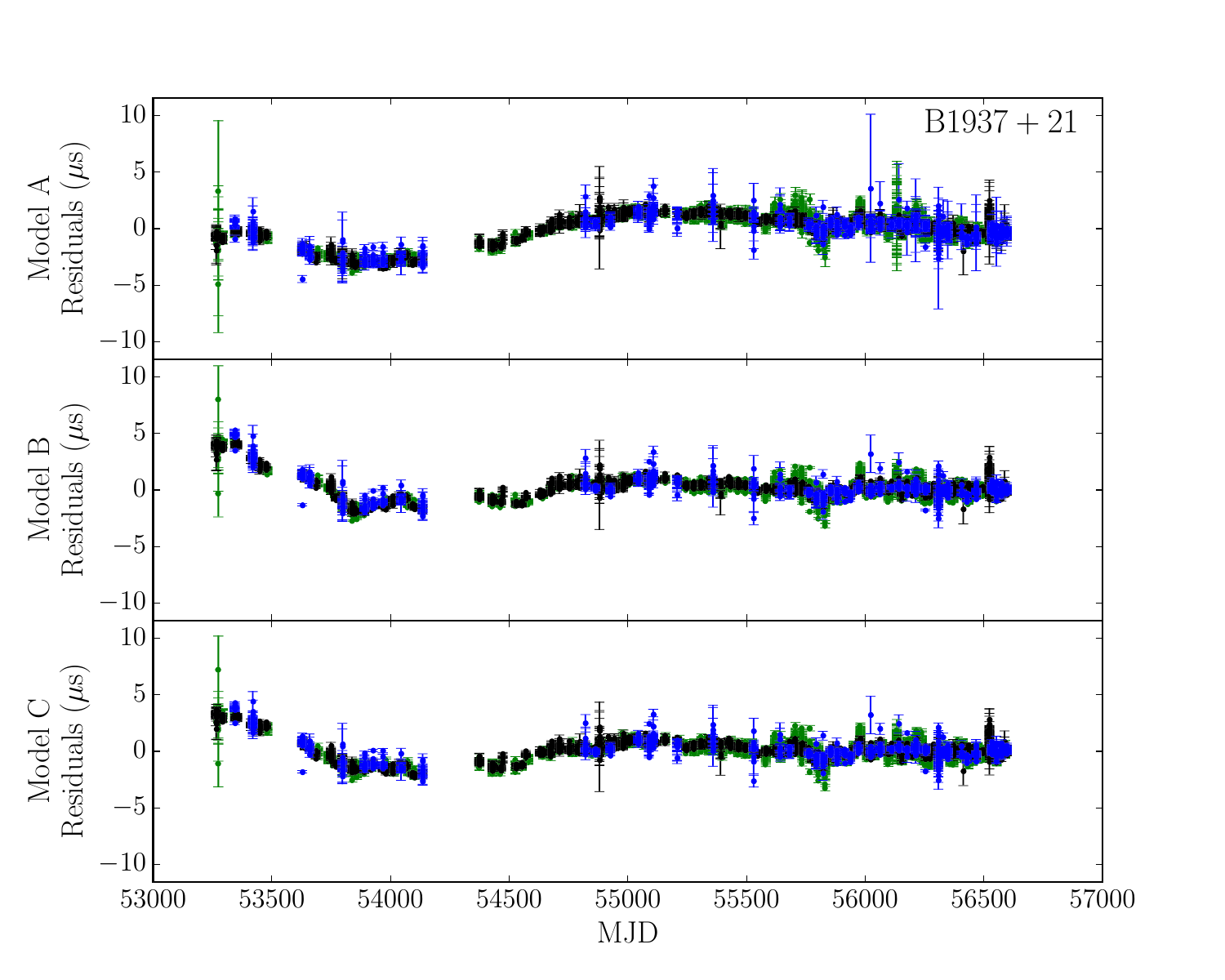}
%\plotone{residcompB1937+21_rasterized.pdf}
  \caption{\footnotesize Comparison of residuals using the different noise models defined in Table~\ref{table:models}. Error bars show the full white-noise TOA uncertainties. Colors indicate different observing bands: 820~MHz (green), 1400~MHz (black), and 2300~MHz (blue).}
\label{fig:B1937+21resids}
\end{center}
\end{figure*}

For convenience, we define the rms noise that is constant with pulse S/N,
\be
\sigma_\const = \sqrt{\sigma_J^2 + \sigma_{\DISS}^2},
\label{eq:sigma_const}
\ee
which we will define with the subscript ``const'' to avoid confusion with the rms of the chromatic excess noise $\sigma_C$.  For many pulsar/backend combinations, NG9WN were able to use $\sigma_\const$ along with scintillation-estimated measurements of $\sigma_{\DISS}$ in order to estimate $\sigma_\J$. For a few pulsar/backend combinations, that was not the case (given the time and frequency resolution of their analysis) and those pulsars will be highlighted later in \S\ref{sec:discussion}.

After we constructed our new TOA errors, we used the \tempotwo~software package accessed via libstempo\footnote{\url{https://github.com/vallis/libstempo}} to refit the timing model \citep{tempo2}. We used the same set of timing parameters as used in NG9 but we refit them after removing the noise model. We used the resulting post-fit residuals (i.e., measured TOAs minus model TOAs) to generate the noise estimates made throughout the present work.  We note that inclusion or exclusion of individual model parameters could bias our results.

Since \tempo~and \tempotwo~will attempt to minimize the $\chi^2$ of the overall timing fit, in some cases they find the minimum by separating TOAs from different receiver bands within a given epoch, producing different sets of residuals as a function of receiver band that do not track each other. The splitting of residuals results from power preferentially being absorbed by the achromatic spin, astrometric, and orbital terms, i.e., there is covariance between the achromatic and chromatic timing parameters. We do not believe these solutions to be physical and therefore we decided to iteratively solve for the best-fit timing solution, facilitated by our use of libstempo.

Starting with the NG9 timing parameter values, we fixed trial values of $\Delta \DM_i$ and $\FD$, refit the achromatic parameters and generated residuals with libstempo, and then determined a new set of fixed trial chromatic parameters using a least-squares method that minimized the ``intra-epoch'' variance, defined as follows. For all epochs defined by the DMX time bins from NG9, we calculated the epoch-averaged residuals $\bm{\widebar{\R}}$ (in matrix form) from the raw residuals $\bm{\R}$ by following the procedure in \citealt{NG5BWM},
\be
\bm{\widebar{\R}} = \left(\mathbf{U}^T\mathbf{C}^{-1}\mathbf{U}\right)^{-1}\mathbf{U}^T\mathbf{C}^{-1}\bm{\R}.
\ee
Here, $\mathbf{U}$ is the ``exploder'' matrix that maps the epochs to the full set of TOAs (see also \citealt{NG5CW}) and the noise covariance matrix can be written as
\ba
\mathbf{C}_{\nu\nu',t t'} &=& \delta_{t t'}\left[\delta_{\nu\nu'}\sigma_{\SN}^2(S) + \sigma_\J^2(\nuband,\tobs)\right] \nonumber \\
&&+ \rho_{\DISS,\nu\nu',t t'} \sigma_{\DISS}^2(\nuband,\Bchan,\tobs)
\ea
following the form in NG9WN, where $\rho_{\DISS}$ encapsulates the frequency and time correlation scales for DISS. The intra-epoch variance is then simply
\be
\sum_i (\R_{ij} - \widebar{\R}_i)^2.
\ee
Note that in practice there can be multiple observations at the same frequency spaced by several days within any given epoch but we subtract a single value from all of the residuals in that time range. Our procedure serves to keep infinite-frequency residuals from the same epoch close in pulse phase while allowing for \tempotwo~to find the best-fit achromatic timing parameters. We found that the $\Delta \DM_i$ parameters typically differed by $\lesssim 1~\sigma$ from the NG9 values except for a few epochs across all pulsars. The $\FD$ parameters typically differed by $\ll 1~\sigma$.

Table~\ref{table:models} provides a summary of three different possible noise models for the timing fit. Model A comes directly from NG9. Model B removes all additional noise parameters except for the template-fitting errors on the TOAs; we show it for comparison purposes only. Again, in order to achieve a phase-connected solution between receiver bands, we use our iterative timing model approach discussed previously. Model C starts with the Model B TOAs and adds in the appropriately-scaled jitter and scintillation noise parameters from NG9WN. It is Model C we test throughout this work.

\begin{figure}
\includegraphics[width=0.5\textwidth]{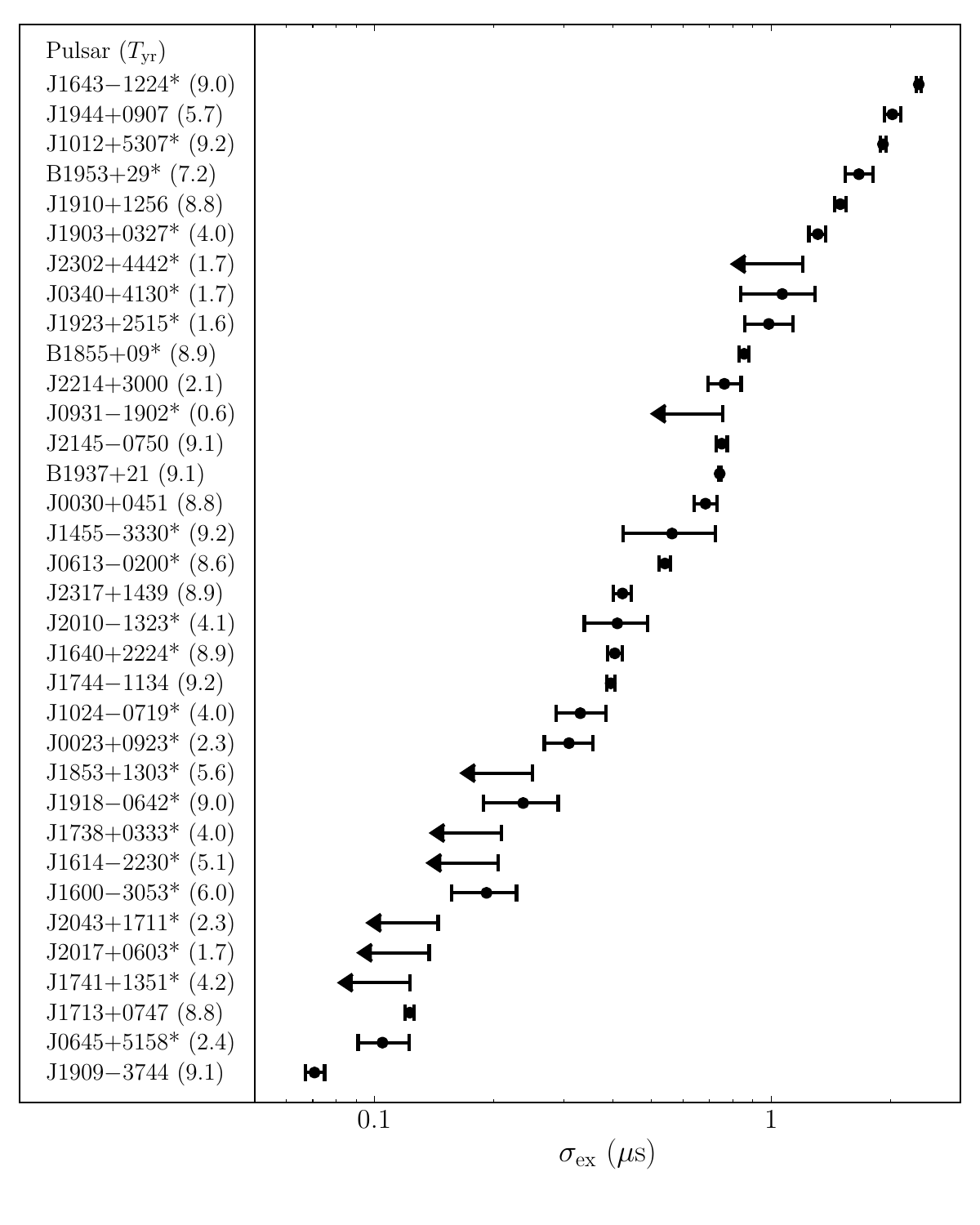}
  \caption{\footnotesize Measured excess noise for each of the NANOGrav pulsars ranked by decreasing value. The values in parentheses show the total timespan observed for each pulsar, in years. Asterisks indicate pulsars for which the white noise model is an upper limit for at least one receiver band. We report detections with the median likelihood value and the $\pm34.1\%$ errors, with upper limits are shown at the 95\% level.}
\label{fig:rednoisesummary}
\end{figure}

\begin{figure*}[t!]
\epsscale{1.2}
\plotone{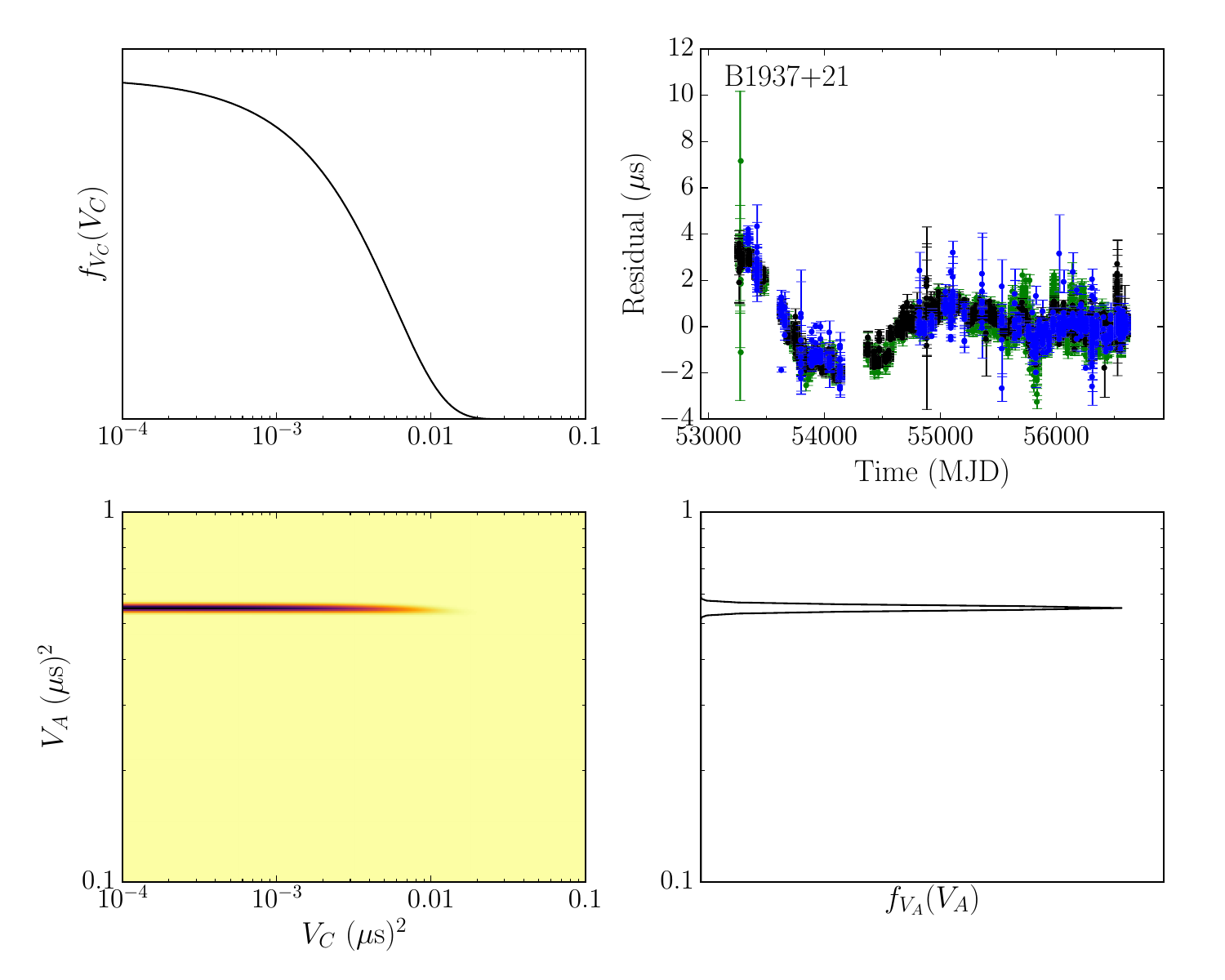}
  \caption{\footnotesize Achromatic/chromatic stochastic process likelihood analysis for PSR~B1937+21. Bottom left: Two-dimensional likelihood surface. The marginalized distributions are shown to the top and right. Top right: timing residuals (820~MHz, green; 1400~MHz, black; 2300~MHz, blue) for reference.}
\label{fig:B1937+21_spin_ism}
\end{figure*}

We show a comparison of the timing residuals for PSR~B1937+21 obtained from the three noise models in Figure~\ref{fig:B1937+21resids}. PSR~B1937+21 has the smallest template-fitting and jitter errors in NG9 within each given receiver band both at AO and GBT (NG9WN) while its scintillation noise is the largest of the three WN terms, which is atypical for pulsars in our data set. We show its residuals as an example to demonstrate the effect of the timing model refit because it is not dominated by template fitting errors and has known red noise residuals \citep[e.g.,][]{sc2010,NG9yr,Lentati+2016}. As expected visually from Figure~\ref{fig:B1937+21resids}, we see that PSR~B1937+21 contains excess noise beyond the WN model. 

Throughout this section, we report detections on the rms excess noise with the median value from the PDFs and the $\pm34.1\%$ errors. Since we perform our likelihood calculations in the logarithm of variance and convert the PDFs to linear rms, we see tails in the distribution of log-variance towards smaller values ($V \to 0$) that become narrow peaks of probability in linear-rms space at $\sigma \approx 0$. Choosing the median value is robust against biasing our results to zero. We find that the median values are within the errors of the maximum likelihood values in log-variance. Detections must be ``$2\sigma$'' significant otherwise we report the non-detection values as 95\% upper limits. The results for our two variance analyses will be reported further down in Table~\ref{table:summary}.

\subsection{Single Stochastic Process}

Using the Model C residuals, we followed the methodology laid out in \S\ref{sec:excess_variance_methodology} for calculating the total excess noise. Figure~\ref{fig:rednoisesummary} shows the results of the likelihood analysis for a single excess noise process (the measured excess beyond the white noise) in the NANOGrav pulsars using the Model C residuals. All values are shown unnormalized to the length of the data span; we expect longer data sets to show increased excess noise, either from the presence of red noise or potentially from increased probability of other noise sources (e.g., RFI) being introduced. The total timespan observed for each pulsar is shown in parentheses.

PSRs~J1909$-$3744 and J1713+0747 are two of the best-timed pulsars observed by NANOGrav, both high in S/N with very low WN (NG9WN). We note that they also show the lowest amounts of detected excess noise of all of the NANOGrav pulsars in our analysis. Many of the pulsars with the highest measured values are consistent with those that displayed ``significant'' red noise in NG9. Five of the top six detections (PSRs~J1643$-$1224, J1944+0907, B1953+29, J1910+1256, and J1903+0327) were the pulsars in NG9 for which unmodeled ISM propagation effects were claimed to be the dominant cause of the excess noise. Our next analysis is better suited for determining if such chromatic effects may be present in the data. The third pulsar, PSR~J1012+5307, also showed measurable red noise in NG9 though without frequency dependence.

\subsection{Achromatic and Chromatic Stochastic Processes}

Again using the Model C residuals, we calculated the variance of an achromatic and a chromatic process in the timing residuals for each pulsar. For the chromatic component, we fixed $\s = 2$ to account for effects such as imperfect DM estimation and define the reference frequency to be $\nu_0 = 1000~$MHz. We show the result of the analysis for PSR~B1937+21 in Figure~\ref{fig:B1937+21_spin_ism}. The timing residuals are shown in the top right with the different receiver bands colored (820~MHz, green; 1400~MHz, black; 2300~MHz, blue). The two-dimensional likelihood surface $\Like(V_A,V_C)$ is shown in the bottom left. The marginalized PDFs are shown to above and to the right of the likelihood. For PSR~B1937+21, we see a well-constrained measurement of achromatic excess noise (AEN) variance but an upper limit on the chromatic excess noise (CEN) variance, consistent with the fact that the timing residuals are correlated in time but the residuals track each other in frequency.

\begin{figure}[t!]
\includegraphics[width=0.5\textwidth]{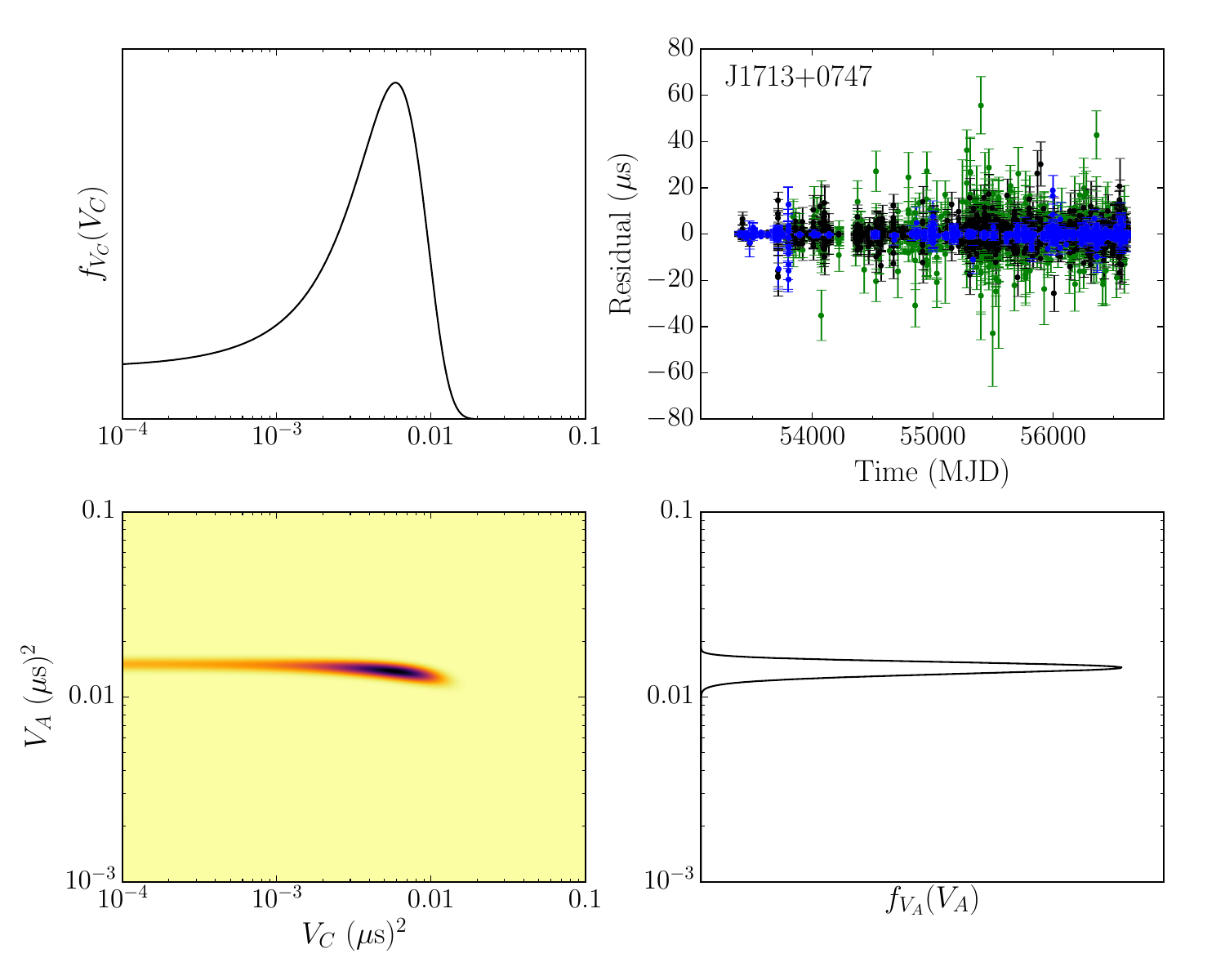}
  \caption{\footnotesize Achromatic/chromatic stochastic process likelihood analysis for PSR~J1713+0747. See the caption to Figure~\ref{fig:B1937+21_spin_ism} for more details.}
\label{fig:J1713+0747_spin_ism}
\end{figure}

\begin{figure}%[t!]
\includegraphics[width=0.5\textwidth]{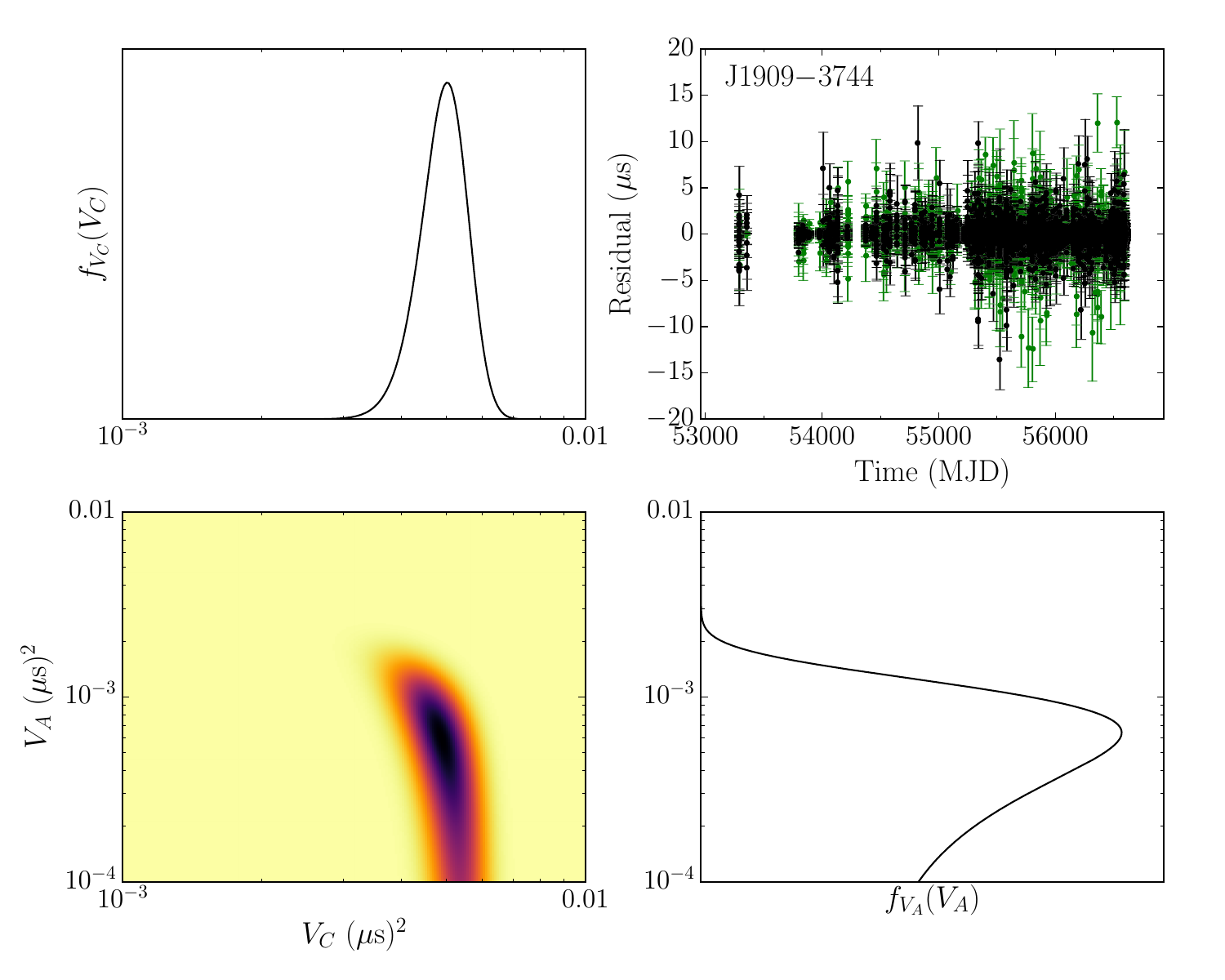}
  \caption{\footnotesize Achromatic/chromatic stochastic process likelihood analysis for PSR~J1909$-$3744. See the caption to Figure~\ref{fig:B1937+21_spin_ism} for more details.}
\label{fig:J1909-3744_spin_ism}
\end{figure}

\begin{figure}%[t!]
\includegraphics[width=0.5\textwidth]{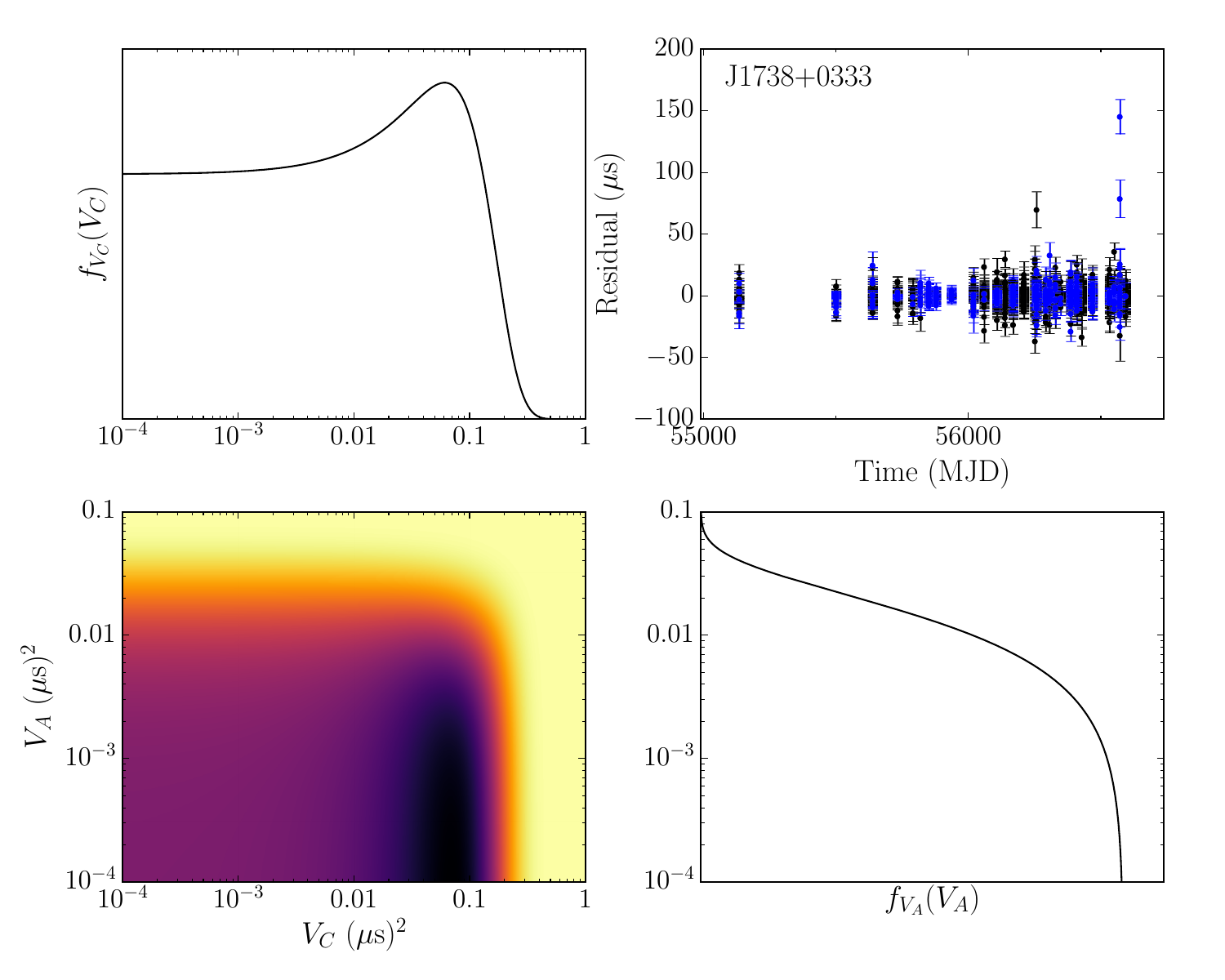}
  \caption{\footnotesize Achromatic/chromatic stochastic process likelihood analysis for PSR~J1738+0333. See the caption to Figure~\ref{fig:B1937+21_spin_ism} for more details.}
\label{fig:J1738+0333_spin_ism}
\end{figure}

\begin{figure}%[t!]
\includegraphics[width=0.5\textwidth]{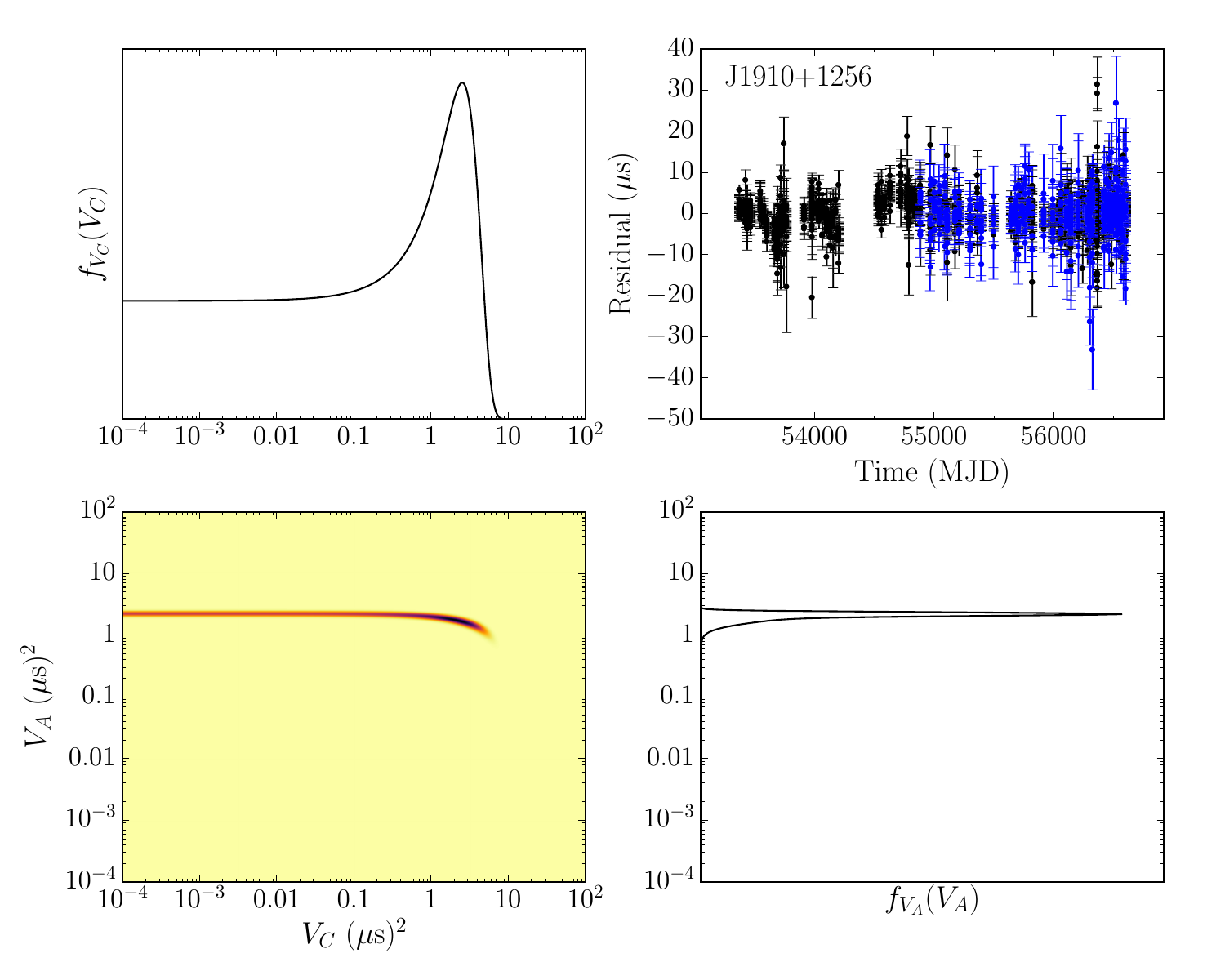}
  \caption{\footnotesize Achromatic/chromatic stochastic process likelihood analysis for PSR~J1910+1256. See the caption to Figure~\ref{fig:B1937+21_spin_ism} for more details.}
\label{fig:J1910+1256_spin_ism}
\end{figure}

Figures~\ref{fig:J1713+0747_spin_ism}$-$\ref{fig:J1910+1256_spin_ism} show the same likelihood analysis for four other pulsars. PSRs J1713+0747 and J1909$-$3744 are two of the best-timed pulsars in the NANOGrav pulsar timing array, each with detections of the variance of both noise processes. The analysis for PSR J1738+0333 is shown as an example of upper limits in both the achromatic and the chromatic excess noise. Lastly, we show the results for PSR J1910+1256 as it has a long stretch of single-frequency observations; therefore some chromatic DM variations may be best described by an achromatic noise process in our likelihood analysis. We discuss the problem in greater detail in the next section.

\begin{figure}%[t!]
\includegraphics[width=0.5\textwidth]{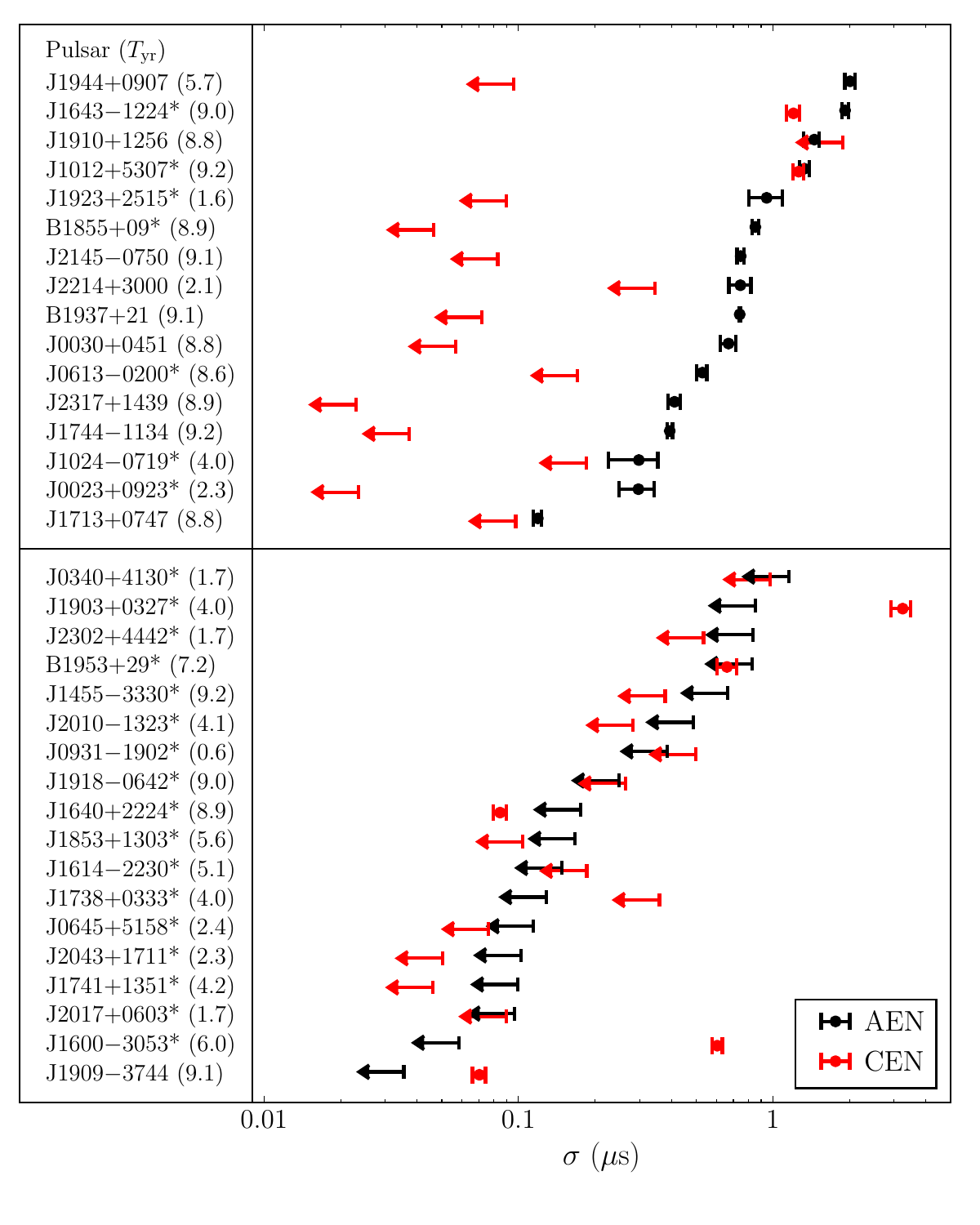}
  \caption{\footnotesize Measured achromatic (AEN, black) and chromatic excess noise (CEN, red) for each of the NANOGrav pulsars ranked by the achromatic noise. The numbers in parentheses are the total timespan observed for each pulsar, in years. Asterisks indicate pulsars for which the white noise model is an upper limit for at least one receiver band. Here we separate the detections of achromatic noise (top) from the upper limits (bottom).}
\label{fig:acsummary}
\end{figure}

Figure~\ref{fig:acsummary} shows the results of our achromatic/chromatic likelihood analyses for all pulsars similar to Figure~\ref{fig:rednoisesummary} in the case of single excess noise process. Note that while we previously plotted the achromatic/chromatic likelihood PDFs in the logarithm of the variance, we calculated the median likelihood values, parameter errors, and upper limits using the marginalized PDFs of the rms achromatic/chromatic errors. In Figure~\ref{fig:acsummary}, we order the pulsars by the AEN to rank by the known irreducible noise, whereas it is unclear if CEN can be corrected for in the future. We re-emphasize that the noise values we report are unnormalized to the length of the total data span; we expect that pulsars observed for longer will in general show increased red, and therefore excess, noise. Again, the total timespan observed is shown in parentheses for each pulsar.

\begin{figure}%[t!]
\includegraphics[width=0.5\textwidth]{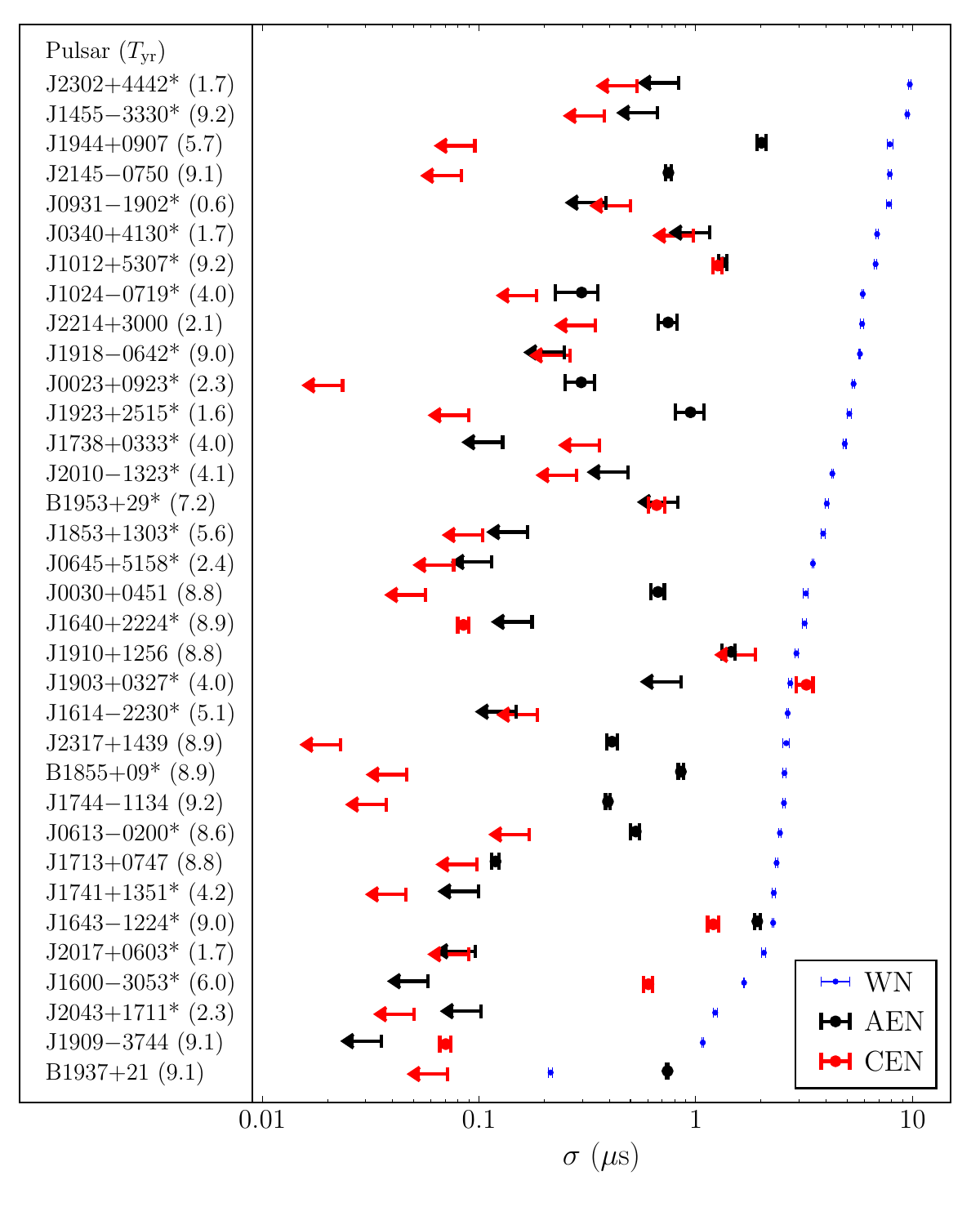}
  \caption{\footnotesize Measured achromatic excess noise (AEN), chromatic excess noise (CEN, red), and white noise (WN, blue) for each of the NANOGrav pulsars ranked by the white noise, showing the total noise budget for the pulsar timing array.  The numbers in parentheses are the total timespan observed for each pulsar, in years. Asterisks indicate pulsars for which the white noise model is an upper limit for at least one receiver band. }
\label{fig:acwsummary}
\end{figure}

In Figure~\ref{fig:acwsummary}, we include estimates of the white noise,
\be
V_W = \frac{1}{N} \sum_{i,j} W_{ij}^2
\label{eq:whitenoisevariance}
\ee
for each pulsar to show the total noise budget for the NANOGrav pulsar timing array. We order the pulsars by their white noise values. We note that the rms WN will contain varying contributions from residuals at each observing band, whose noise properties change as a function of receiver band. Changes in the length of each individual observation will also affect the TOA errors and therefore the measured white noise. As a result, even one of our most precisely-timed pulsars like PSR~J1713+0747 can be ranked higher up in the plot. We emphasize that we are measuring the total sum of the white noise in the nine-year data release, and that future predictions of white-noise performance should come from NG9WN.

\begin{figure}[t!]
\hspace{-2ex}
\includegraphics[width=0.5\textwidth]{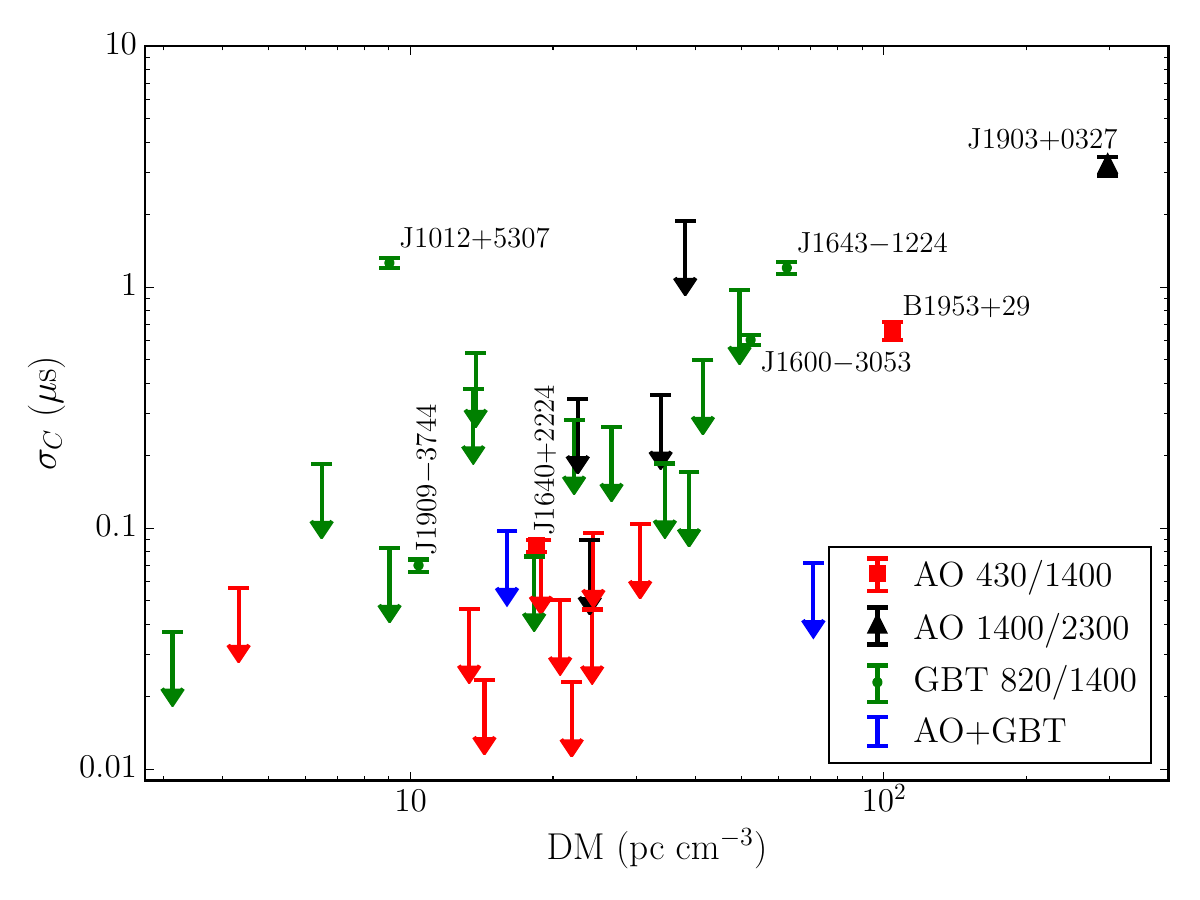}
  \caption{\footnotesize Chromatic excess noise (CEN) in NANOGrav pulsars versus DM. Colors denote which telescope and receiver bands were used to observe a pulsar. The two blue points represent PSRs J1713+0747 (820/1400 at GBT and 1400/2300 at AO) and B1937+21 (820/1400 at GBT and 430/1400 at AO).}
\label{fig:red_vs_DM}
\end{figure}

Of the five pulsars in NG9 that were claimed to have red noise likely from ISM propagation effects, three show CEN in our analysis: PSRs J1643$-$1224, J1903+0327, and B1953+29. These three pulsars are three of the top four highest DM pulsars in NG9. In Figure~\ref{fig:red_vs_DM}, we explore the possible relationship between $\sigma_C$ and ISM noise by plotting versus DM. Colors show which telescopes and receiver bands were used to observe each pulsar. While generally the CEN increases with DM, PSR~J1012+5307 is an outlier. Frequency dependence between the two bands of residuals for PSR~J1012+5307 is clear for several epochs upon visual inspection but the cause, astrophysical or otherwise, is presently unknown.

\begin{figure}[t!]
\hspace{-8ex}
\includegraphics[width=0.6\textwidth]{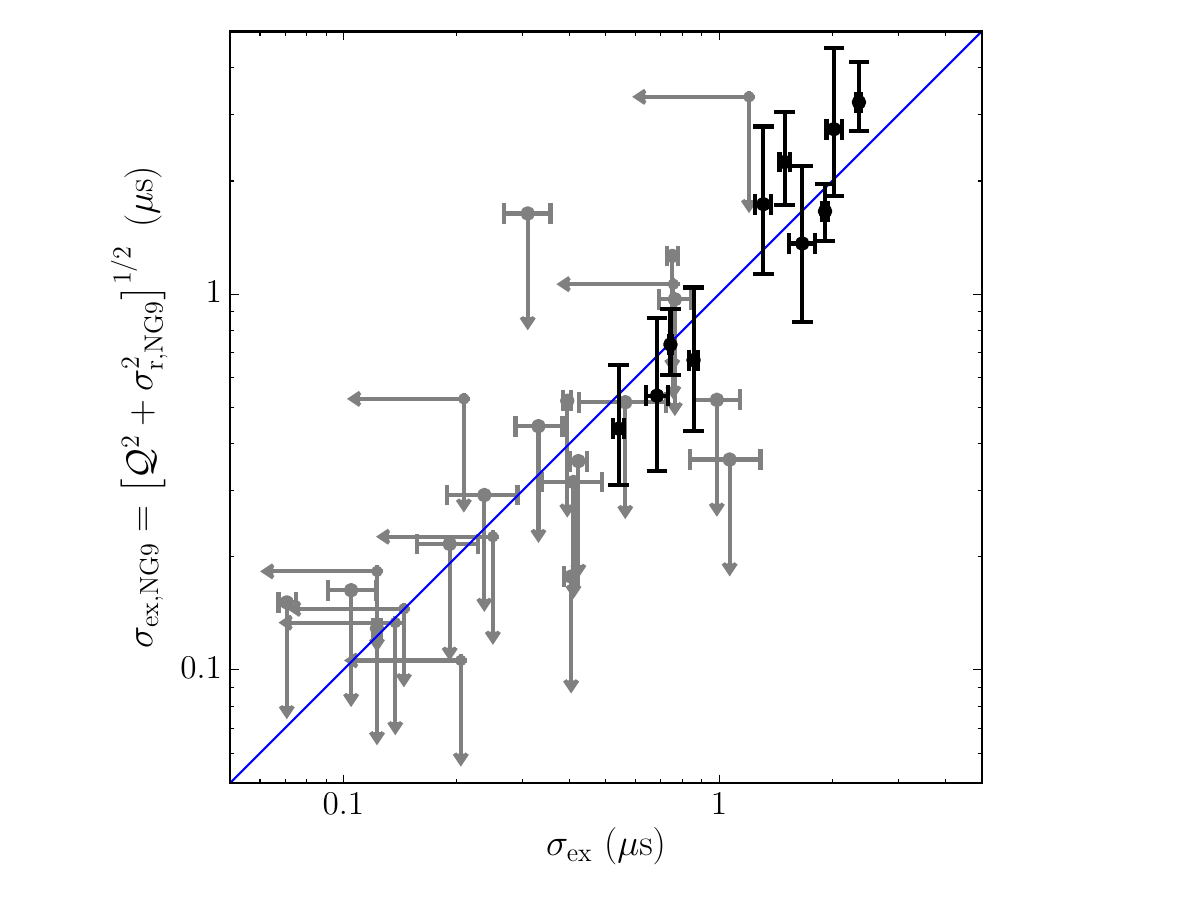}
  \caption{\footnotesize Comparison of excess noise between NG9 and this work. Black points denote measurements in both while the gray points mark upper limits in at least one of the values for a given pulsar. The diagonal blue line shows where the NG9 excess noise equals $\sigex$.}
\label{fig:comparison}
\end{figure}

\subsection{Comparison with the NG9 Noise Parameters}

We compare our single-process excess noise measurements with the estimated noise from NG9. We choose the single-process parameterization as a more equal comparison with the previous results because chromaticity is not accounted for in NG9. Our excess noise measurements correspond to the quadratic sum of the red noise and $\mathcal{Q}$ (EQUAD).

Following the formalism in NG9, we can convert the red noise parameters $\Ared$ and $\gammared$ into an rms ``timing noise'' after a quadratic fit of the residuals, which we can write in the form\footnote{We note that Eq.~\ref{eq:sigmaTN2} corrects two typos in Eq.~6 in NG9, the factor of ``$(1-\gammared)$'' should be ``$-(1+\gammared)$'' when the spectral index is defined oppositely as it is here, i.e. $P(f) \propto f^\gammared$. Also, the pre-factor of 2.05 should be 3.0 ns. }
\ba
\sigma_{\rm TN,2} & \approx & \left(\int_{1/T}^\infty P(f)df\right)^{1/2} \nonumber \\
&=& 3~\mathrm{ns}~\left(\gammared-1\right)^{-1/2} \times \nonumber \\
&& \left(\frac{\Ared}{3\times 10^{-3}~\mathrm{\mu s~yr^{1/2}}}\right) T_{\mathrm{yr}}^{(\gammared-1)/2}.
\label{eq:sigmaTN2}
\ea
The result is only valid for power-law spectra when $\gammared > 1$. When $\gammared = 1$, the variance is proportional to $\ln(f_h T)$, where $f_h$ is a high-frequency cutoff roughly set by the observing cadence. When $\gammared < 1$, the variance is proportional to $f_h^{(1-\gammared)}$ and does not depend on $T$. The use of $1/T$ is Eq.~\ref{eq:sigmaTN2} as the lower-cutoff frequency is only an approximation of a filter for quadratic subtraction, the fitting for spin period and spin period derivative in the pulsar timing model. We used a more exact approach in calculating the post-timing-model-fit rms that included a normalized transmission function $\mathcal{T}(f)$ derived from the spin and astrometric timing parameters that NG9 fit to all pulsars \citep{brn1984}. The post-fit rms is therefore smaller than the value given by Eq.~\ref{eq:sigmaTN2}. Using the formalism in \citet{mcc2013}, we calculated $\mathcal{T}(f)$ and applied it to the power-law power spectra for each pulsar to numerically determine the post-fit rms from red noise as
\be
\sigma_{\red,\rm post} = \left(\int_{1/T}^\infty \mathcal{T}(f)P(f)df\right)^{1/2}.
\label{eq:postfit}
\ee
For the white noise component, since NG9 provided one value of $\mathcal{Q}$ per backend-receiver combination, we calculated the total contribution to the excess noise from $\mathcal{Q}$ as the square root of the average quadrature value from each band weighted by the number of TOAs for each backend-receiver combination. 

Figure~\ref{fig:comparison} shows the comparison between the excess noise estimated in NG9, $\sigma_{\rm ex,NG9} = \left[\mathcal{Q}^2 + \sigma_{\red,\rm NG9}^2\right]^{1/2}$, and $\sigex$, where measurements in both are given by the black points and gray points denote the cases where there is an upper limit in at least one of the two methods. In general, our measurements are consistent with those in NG9 although with much smaller uncertainties (again because of the large covariances between the NG9 red noise parameters); there are a few exceptions in the upper limits.

\subsection{Structure Function Analysis}
\label{sec:SF}

Up to this point, we have ignored the spectral properties of the excess noise, i.e., we have not attempted to decompose the excess noise into red and white components. Residuals with red noise will have increased power at lower fluctuation frequencies in their ensemble-average power spectrum. We can use a structure function (SF) analysis to show that there is excess power at large time lags (low fluctuation frequencies) and demonstrate the contribution of red noise to the total excess noise budget. We expect some pulsars to be dominated by red noise upon visual inspection of the residuals, such as for PSR~B1937+21. Individual pulsars with other noise properties will be discussed in the next section.  An analysis of additional properties of the power spectra are outside the scope of this paper.

The SF for a time series $X(t)$ is the average squared difference/increment
\be
D_X(\tau) = \left<\left[X(t+\tau)-X(t)\right]^2\right>,
\label{eq:SF}
\ee
where $\tau$ is a time lag. It is trivial to show that for white noise $W(t)$ with variance $V_W$, the SF is $2V_W(1-\delta_{\tau 0})$, where $\delta$ is the Kronecker delta. Generally, the SF will be related to the variance at a given non-zero lag by a factor of 2.

Within a given epoch, we have many residuals measured at different observing frequencies that contribute large amounts of WN to the SF. Therefore, in order to reduce the influence of the WN, we computed epoch-averaged residuals as we discussed in \S\ref{sec:new_timing_model}. Epoch-averaging reduces the WN variance roughly by a factor of the number of residuals in an epoch (assuming equal errors) whereas the red noise variance will remain constant.

\begin{figure}[t!]
\includegraphics[width=0.5\textwidth]{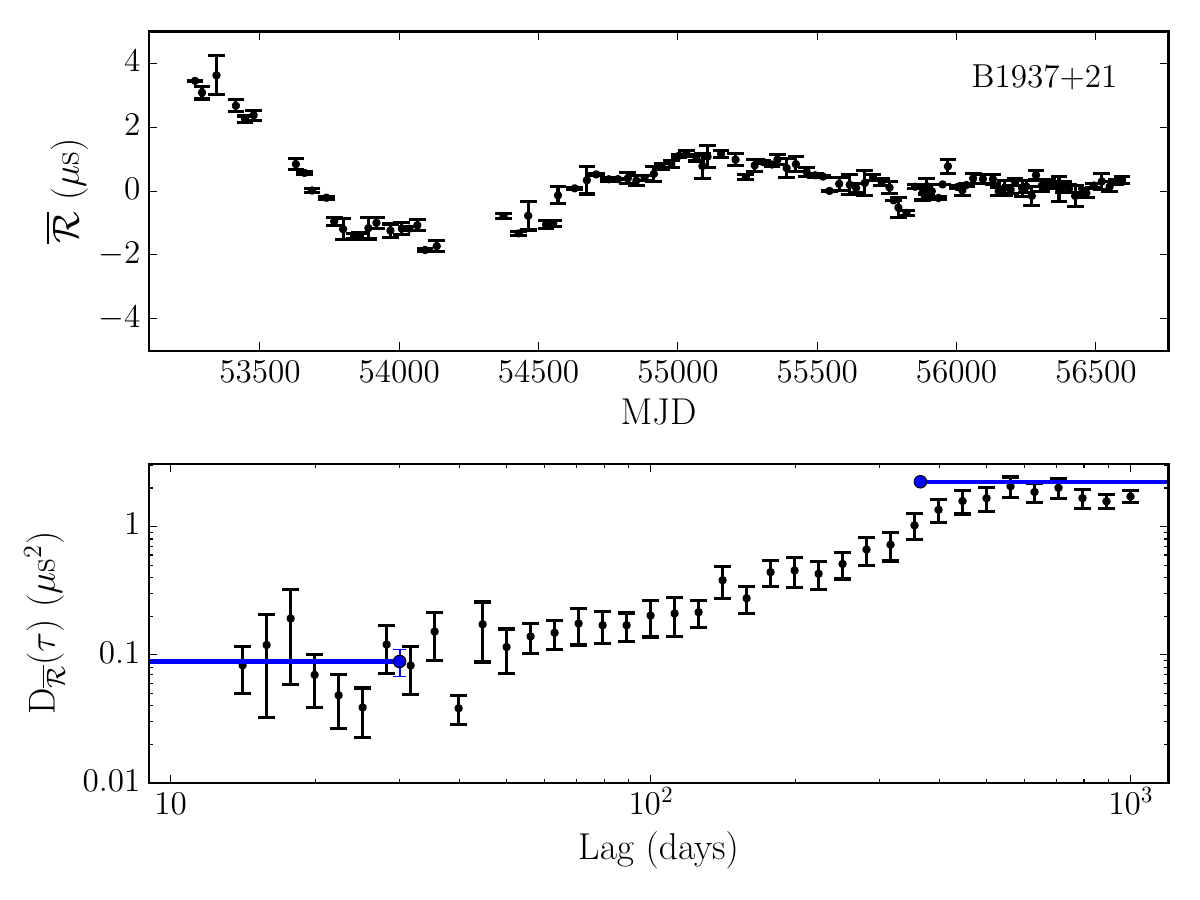}
  \caption{\footnotesize Epoch-averaged residuals (top) and structure function (bottom) for PSR~B1937+21 with 20 lag bins per decade. The blue lines indicate the two lag bin ranges defined as $0 < \tau \leq 30$~days and $\tau \geq 365$~days.}
\label{fig:sf_B1937+21}
\end{figure}

\begin{figure}[t!]
\includegraphics[width=0.5\textwidth]{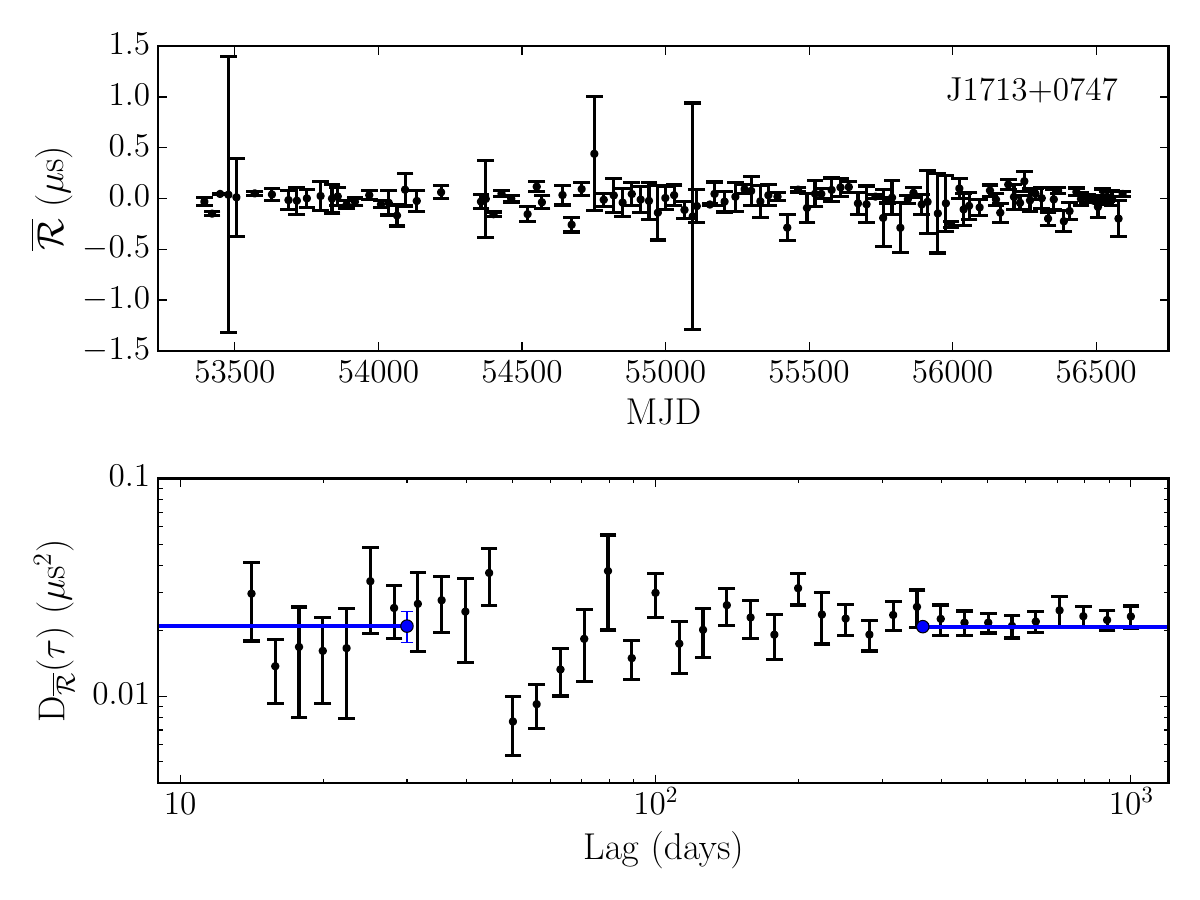}
  \caption{\footnotesize Epoch-averaged residuals (top) and structure function (bottom) for PSR~J1713+0747 with 20 lag bins per decade. The blue lines indicate the two lag bin ranges defined as $0 < \tau \leq 30$~days and $\tau \geq 365$~days.}
\label{fig:sf_J1713+0747}
\end{figure}

Figures~\ref{fig:sf_B1937+21} and \ref{fig:sf_J1713+0747} show the epoch-averaged residuals (denoted as $\overline{\R}$) and the SFs for PSRs~B1937+21 and J1713+0747 respectively. We used 20 lag bins per decade and the maximum lag bin shown is at $\tau = 1000$ days; small number statistics cause wild fluctuations for narrowly-defined large lag bins and so we do not show them. Errors are taken to be the rms of the increments in each lag bin. We find that many of our pulsars show increases in the SF at large ($\tau \geq 365$~days) lags over small ($0 < \tau \leq 30$~days, up to the typical cadence for each pulsar at each telescope) lags; the blue points/lines show the SF for the bins defined over the two broad ranges. We ignored contributions from the zero lag because the increments are zero, which will affect the mean of the squared increments for lags $\leq 30$~days. Increases in the SF indicate that red noise is present in a number of the pulsars and in some cases dominates over temporally-uncorrelated excess noise; for pure white noise the SF values would be comparable.

To estimate the red variance, we define the redness parameter to be the ratio of the SF over the two broad lag bins,
\ba
\RP &\equiv& \frac{D_{\overline{\R}}(\tau \geq 365)}{D_{\overline{\R}}(0<\tau \leq 30)} \nonumber \\
&\approx& \frac{2V_r + 2V_W}{2V_W} \approx \frac{V_r}{V_W} + 1.
\ea
Our redness parameter is similar in spirit to the one defined by \citet{cd1985} which characterizes the redness of the power spectra using the slope of the third-order SF of the (phase) residuals. The parameter is $\RP_{\rm CD} = (m_3 -1)/2$, where $m_3$ is the slope of the third-order SF. The data used here are too sparse to properly estimate higher-order SFs. First-order SFs like in Eq.~\ref{eq:SF} scale as $\tau^{\gammared-1}$ for $\gammared \leq 3$ and saturate as $\tau^2$ for steeper spectra; higher-order SFs allow for the characterization of steeper slopes (see e.g., \citealt{fra2010}). Rather than impose a functional form of the power spectra in this work, we used the first-order SF to simply characterize the bulk power at low versus high fluctuation frequencies (large versus small time lags), parameterized by our form of $\RP$. 

Along with the results of the variance analyses, we show the values of $\RP$ for each pulsar in Table~\ref{table:summary}. Errors reflect the scatter of the increments in the two time lag bins propagated from the SFs into each individual $\RP$ only; we have not accounted for realization errors \citep{Lam+2015}. The uncertainties on the epoch-averaged residuals are not included and contribute non-trivially; however, any increase in the uncertainty of $\RP$ will be minimized by the large number of increments being averaged together. For convenience, we also show the fraction of the red variance versus the total variance, which is simply 
\be
\frac{V_r}{V_{\rm tot}} = \frac{\RP-1}{\RP}
\ee
if we assume that the only contributions to the SF can be broken into red and white noise components. We note that pulsars with $\RP$ values inconsistent with and less than 1 have short total observing timespans, which will cause poor estimation of the large time-lag bin. All other $\RP$ values are consistent with 1 or greater. To test for biases in our metric, we performed simulations in which we generated idealized TOAs from timing models with WN perturbations, refit the timing model to produce residuals, and then calculated the RP. When the total timespan of the simulated residuals was long ($\sim 9$~yr), we found a small (few percent) bias but $\RP \approx 1$, well within the measurement errors for the real data. Thus we believe that significant departures from 1 demonstrate the presence of red noise, which we see in several of our pulsars.

\begin{deluxetable*}{lccc|c|cc|c|cc}
\tablecolumns{10}
\tablecaption{Summary of Excess Noise Parameters}
\tablehead{
\colhead{Pulsar} & \colhead{Obs} & \colhead{DM} & \colhead{$T$} & \colhead{$\sigma_{\mathrm{ex}}$} & \colhead{$\sigma_{A}$} & \colhead{$\sigma_{C}~\mathrm{at}~1~\mathrm{GHz}$} & \colhead{$\sigma_{W}$} & \colhead{$\RP$} & \colhead{$V_\red/V_{\rm tot}$}\\
\colhead{} & \colhead{} & \colhead{(pc cm$^{-3}$)}  & \colhead{(yr)}  & \colhead{$\mathrm{(\mu s)}$} & \colhead{$\mathrm{(\mu s)}$} & \colhead{$\mathrm{(\mu s)}$} & \colhead{$\mathrm{(\mu s)}$} & \colhead{} & \colhead{}
}
\startdata
J0023+0923* & AO & 14.3 & 2.3 & $0.31_{-0.04}^{+0.05}$ & $0.30_{-0.05}^{+0.04}$ & $<\!0.02$ & $5.35 \pm 0.08$ & $0.5 \pm 0.2$ & $-1.04 \pm 0.87$\\*
J0030+0451 & AO & 4.3 & 8.8 & $0.68_{-0.04}^{+0.05}$ & $0.67 \pm 0.05$ & $<\!0.06$ & $3.22 \pm 0.08$ & $9.0 \pm 2.5$ & $0.89 \pm 0.03$\\*
J0340+4130* & GBT & 49.6 & 1.7 & $1.1 \pm 0.2$ & $<\!1.16$ & $<\!0.98$ & $6.87 \pm 0.06$ & $1.9 \pm 0.7$ & $0.48 \pm 0.20$\\*
J0613$-$0200* & GBT & 38.8 & 8.6 & $0.54 \pm 0.02$ & $0.53_{-0.03}^{+0.02}$ & $<\!0.17$ & $2.45 \pm 0.04$ & $9.3 \pm 2.4$ & $0.89 \pm 0.03$\\*
J0645+5158* & GBT & 18.2 & 2.4 & $0.10_{-0.01}^{+0.02}$ & $<\!0.11$ & $<\!0.08$ & $3.47 \pm 0.04$ & $0.6 \pm 0.2$ & $-0.64 \pm 0.67$\\*
J0931$-$1902* & GBT & 41.5 & 0.6 & $<\!0.76$ & $<\!0.39$ & $<\!0.50$ & $7.80 \pm 0.21$ & $-$ & $-$\\*
J1012+5307* & GBT & 9.0 & 9.2 & $1.91 \pm 0.03$ & $1.33 \pm 0.06$ & $1.26 \pm 0.06$ & $6.76 \pm 0.08$ & $1.4 \pm 0.2$ & $0.28 \pm 0.12$\\*
J1024$-$0719* & GBT & 6.5 & 4.0 & $0.33_{-0.04}^{+0.05}$ & $0.30_{-0.07}^{+0.06}$ & $<\!0.18$ & $5.90 \pm 0.07$ & $1.5 \pm 0.3$ & $0.34 \pm 0.14$\\*
J1455$-$3330* & GBT & 13.6 & 9.2 & $0.6_{-0.1}^{+0.2}$ & $<\!0.67$ & $<\!0.38$ & $9.47 \pm 0.11$ & $1.8 \pm 0.4$ & $0.43 \pm 0.14$\\*
J1600$-$3053$^a$* & GBT & 52.3 & 6.0 & $0.19_{-0.03}^{+0.04}$ & $<\!0.06$ & $0.61 \pm 0.03$ & $1.67 \pm 0.01$ & $1.7 \pm 0.3$ & $0.42 \pm 0.11$\\*
J1614$-$2230* & GBT & 34.5 & 5.1 & $<\!0.21$ & $<\!0.15$ & $<\!0.19$ & $2.66 \pm 0.02$ & $4.0 \pm 1.6$ & $0.75 \pm 0.10$\\*
J1640+2224$^a$* & AO & 18.5 & 8.9 & $0.40 \pm 0.02$ & $<\!0.18$ & $0.085 \pm 0.005$ & $3.18 \pm 0.07$ & $2.6 \pm 0.7$ & $0.61 \pm 0.10$\\*
J1643$-$1224$^a$* & GBT & 62.4 & 9.0 & $2.36_{-0.03}^{+0.04}$ & $1.92 \pm 0.06$ & $1.21 \pm 0.07$ & $2.28 \pm 0.02$ & $2.2 \pm 0.7$ & $0.55 \pm 0.14$\\*
J1713+0747 & AO+GBT & 16.0 & 8.8 & $0.123 \pm 0.003$ & $0.119_{-0.005}^{+0.004}$ & $<\!0.10$ & $2.36 \pm 0.04$ & $1.0 \pm 0.2$ & $-0.01 \pm 0.16$\\*
 & AO &  & 8.8 & $0.116_{-0.003}^{+0.004}$ & $0.115\pm 0.004$ & $< 0.05$ & $1.23 \pm 0.04$ &  & \\*
 & GBT &  & 7.7 & $0.123_{-0.006}^{+0.007}$ & $0.132 \pm 0.009$ & $0.09_{-0.04}^{+0.02}$ & $2.72 \pm 0.04$ &  & \\*
J1738+0333* & AO & 33.8 & 4.0 & $<\!0.21$ & $<\!0.13$ & $<\!0.36$ & $4.87 \pm 0.07$ & $1.0 \pm 0.9$ & $-0.04 \pm 1.01$\\*
J1741+1351* & AO & 24.2 & 4.2 & $<\!0.12$ & $<\!0.10$ & $<\!0.05$ & $2.29 \pm 0.05$ & $0.4 \pm 0.1$ & $-1.78 \pm 1.16$\\*
J1744$-$1134 & GBT & 3.1 & 9.2 & $0.39 \pm 0.01$ & $0.39 \pm 0.01$ & $<\!0.04$ & $2.55 \pm 0.03$ & $0.7 \pm 0.2$ & $-0.37 \pm 0.34$\\*
J1853+1303* & AO & 30.6 & 5.6 & $<\!0.25$ & $<\!0.17$ & $<\!0.10$ & $3.87 \pm 0.08$ & $2.1 \pm 0.7$ & $0.53 \pm 0.15$\\*
B1855+09* & AO & 13.3 & 8.9 & $0.86 \pm 0.02$ & $0.85 \pm 0.02$ & $<\!0.05$ & $2.57 \pm 0.05$ & $2.9 \pm 1.5$ & $0.65 \pm 0.18$\\*
J1903+0327$^a$* & AO & 297.6 & 4.0 & $1.31_{-0.07}^{+0.06}$ & $<\!0.86$ & $3.2_{-0.3}^{+0.2}$ & $2.73 \pm 0.05$ & $4.3 \pm 1.3$ & $0.77 \pm 0.07$\\*
J1909$-$3744 & GBT & 10.4 & 9.1 & $0.071 \pm 0.004$ & $<\!0.04$ & $0.070 \pm 0.004$ & $1.08 \pm 0.01$ & $0.8 \pm 0.1$ & $-0.20 \pm 0.22$\\*
J1910+1256 & AO & 38.1 & 8.8 & $1.49 \pm 0.05$ & $1.46_{-0.14}^{+0.06}$ & $<\!1.88$ & $2.92 \pm 0.05$ & $2.1 \pm 0.5$ & $0.51 \pm 0.11$\\*
J1918$-$0642* & GBT & 26.6 & 9.0 & $0.24 \pm 0.05$ & $<\!0.25$ & $<\!0.26$ & $5.73 \pm 0.05$ & $1.5 \pm 0.3$ & $0.34 \pm 0.14$\\*
J1923+2515* & AO & 18.9 & 1.6 & $1.0_{-0.1}^{+0.2}$ & $0.9 \pm 0.1$ & $<\!0.09$ & $5.11 \pm 0.12$ & $0.6 \pm 0.3$ & $-0.82 \pm 0.86$\\*
B1937+21 & AO+GBT & 71.0 & 9.1 & $0.742 \pm 0.006$ & $0.741 \pm 0.006$ & $<\!0.07$ & $0.21 \pm 0.01$ & $25.2 \pm 5.9$ & $0.960 \pm 0.009$\\*
 & AO &  & 8.9 & $0.703_{-0.010}^{+0.011}$ & $0.702_{-0.010}^{+0.011}$ & $< 0.10$ & $0.18\pm 0.01$ &  & \\*
 & GBT &  & 9.1 & $0.756 \pm 0.007$ & $0.755\pm 0.007$ & $<0.06$ & $0.22\pm 0.01$ &  & \\*
J1944+0907 & AO & 24.3 & 5.7 & $2.02_{-0.09}^{+0.10}$ & $2.01_{-0.09}^{+0.10}$ & $<\!0.10$ & $7.87 \pm 0.24$ & $2.5 \pm 1.5$ & $0.60 \pm 0.24$\\*
B1953+29* & AO & 104.5 & 7.2 & $1.7 \pm 0.1$ & $<\!0.83$ & $0.66 \pm 0.06$ & $4.03 \pm 0.08$ & $1.3 \pm 0.6$ & $0.24 \pm 0.34$\\*
J2010$-$1323* & GBT & 22.2 & 4.1 & $0.41_{-0.07}^{+0.08}$ & $<\!0.49$ & $<\!0.28$ & $4.27 \pm 0.04$ & $1.5 \pm 0.5$ & $0.34 \pm 0.20$\\*
J2017+0603* & AO & 23.9 & 1.7 & $<\!0.14$ & $<\!0.10$ & $<\!0.09$ & $2.06 \pm 0.05$ & $0.4 \pm 0.2$ & $-1.38 \pm 1.25$\\*
J2043+1711* & AO & 20.7 & 2.3 & $<\!0.14$ & $<\!0.10$ & $<\!0.05$ & $1.23 \pm 0.03$ & $0.5 \pm 0.2$ & $-0.96 \pm 0.92$\\*
J2145$-$0750 & GBT & 9.0 & 9.1 & $0.75 \pm 0.02$ & $0.75 \pm 0.02$ & $<\!0.08$ & $7.86 \pm 0.13$ & $1.0 \pm 0.3$ & $0.04 \pm 0.26$\\*
J2214+3000 & AO & 22.6 & 2.1 & $0.76_{-0.07}^{+0.08}$ & $0.75 \pm 0.07$ & $<\!0.34$ & $5.86 \pm 0.09$ & $0.7 \pm 0.2$ & $-0.39 \pm 0.48$\\*
J2302+4442* & GBT & 13.7 & 1.7 & $<\!1.20$ & $<\!0.83$ & $<\!0.53$ & $9.73 \pm 0.11$ & $0.5 \pm 0.2$ & $-1.08 \pm 0.82$\\*
J2317+1439 & AO & 21.9 & 8.9 & $0.42 \pm 0.02$ & $0.41 \pm 0.02$ & $<\!0.02$ & $2.62 \pm 0.09$ & $1.1 \pm 0.5$ & $0.10 \pm 0.41$
\enddata
\label{table:summary}
\tablenotetext{}{Columns are: pulsar name, observatory, dispersion measure (DM), observed time span ($T$), single-process excess noise ($\sigma_{\mathrm{ex}}$), achromatic excess noise ($\sigma_A$), chromatic excess noise ($\sigma_C$), white noise ($\sigma_W$), redness parameter ($\RP = D_{\widebar{\R}}(\tau\geq 365)/D_{\widebar{\R}}(\tau \leq 30)$, excluding PSR~J0931$-$1902 because $T \le 1$~year), and the fraction of red noise variance ($V_\red/V_{\rm tot}$). For PSRs~J1713+0747 and B1937+21, we show the variance analyses after the full timing model fit but separately considering residuals from the individual telescopes.}
\tablenotetext{*}{Denote pulsars where there is an upper limit in the WN parameters for at least one frequency band.}
\tablenotetext{a}{Pulsars where $\sigma_\const$ is smaller than the scintillation-estimated $\sigma_{\DISS}$.}
\end{deluxetable*}

Of our pulsars, 17 show significance in $V_r/V_{\rm tot}$ at the ``$2\sigma$'' level. When combined with our $2\sigma$ criterion for the significance of an excess noise detection versus an upper limit, we find that 15 pulsars display red noise in the single-process excess noise. All pulsars with detected red noise in NG9 also show redness in our analysis.

\section{Individual Pulsar Discussion}
\label{sec:discussion}

In this section, we will describe the individual results for the NANOGrav pulsars and discuss the broader context for each. We begin with individual pulsars in name order, followed by pulsars grouped by specific characteristics at the end.

\subsection{PSR~J0023+0923}

PSR~J0023+0923 shows a detection of AEN after only $\sim$~2~years of observation despite upper limits in the WN model for both receiver bands. We expect strong scintillation events to allow for future estimation of $\sigma_\J$ and therefore refinement of both the WN and AEN.

\subsection{PSRs~J0340+4130} %slight V_C peak, both have upper limits of sigma_J

PSR~J0340+4130 shows an upper limit in both AEN and CEN but the log-variance likelihood for CEN show a strong peak. With a short observed timespan (1.7~years), increased observations should improve the estimate of the CEN value if the noise grows with time. Conservative errors from the upper limit on the WN however cause a bias in the localization of the peak. The bias will improve with better estimates of the WN. Regardless, the large value of $\niss$ measured in NG9WN indicates that achieving a high enough pulse $\SN$ via scintillation to measure $\sigma_\J$ and therefore constrain the WN parameters may be difficult in the future.

\subsection{PSR~J0645+5158} %slight V_C and V_A peak

PSR~J0645+5158 shows strong peaks in the log-variance likelihoods for both AEN and CEN after 2.4~years of observation though we report both as upper limits based on our $2\sigma$ cutoff criterion (the single-process excess is detected). The upper limit on WN for the 1400~MHz band implies our conservative WN will remove variance from both possible red noise processes and therefore better estimation will yield improved localization of the peaks. The pulsar strongly scintillates and therefore with increased data, we believe refinement of all of the noise terms should be possible in the near future.

\subsection{PSR~J1012+5307}

PSR~J1012+5307 shows strong detections of both AEN and CEN after $\sim$9~years of observation with approximately 3/4th of the excess being red. As discussed in \S\ref{sec:excess_variance_estimation}, PSR~J1012+5307 shows a large CEN compared with its DM. It does show an upper limit on the WN and therefore both red components will be biased. The pulsar exhibits strong scintillation events although since it has been observed the full $\sim$9~years we believe it will be difficult to obtain enough high-$\SN$ pulses to determine $\sigma_\J$. There is currently no measurement of the scintillation timescale for this pulsar, indicating it is of order typical observation timescales or longer; the NE2001 model estimates $\Dtd\sim 20$~minutes at 1400~MHz \citep{NE2001}. Therefore, several long observation tracks on the pulsar may help in capturing enough bright scintles for $\sigma_\J$ determination and thus allow for faster WN parameter refinement.

\subsection{PSR~J1024$-$0719}

A number of timing residuals for PSRs J1024$-$0719 deviate significantly (many sigma) from zero and therefore contribute to the noise estimation as an excess noise beyond the WN. The values of AEN are determined to be upper limits regardless, though the true values are likely significantly lower than that reported. The cause of such discrepant residuals should be investigated in future data releases.

Additionally, \citet{Reardon+2016} find that PSR~J1024$-$0719 shows steep red noise, $P(f) \sim f^{-6}$, in disagreement with our findings where we see significantly lower levels of red noise. The pulsar was recently shown to be in a long-period orbit \citep{Kaplan+2016,Bassa+2016}, where differences in the timing models for the pulsar were noted because of the unaccounted for binary. It is possible that differences in the timing models account for the disagreement in the estimated red noise.

\subsection{PSR~J1455$-$3330}

As with PSR~J1024$-$0719, a number of timing residuals for PSR~J1455$-$3330 deviate significantly from zero in a non-stationary way and contribute to an increase in the AEN (though an upper limit was also determined for this pulsar). The cause of the outliers is unknown and should be further investigated.

\subsection{PSR~J1643$-$1224}

PSR~J1643$-$1224 has demonstrated unusual chromatic timing behavior \citep{Perrodin+2013,NG5BWM,Lentati+2016}; pulse shape variations are possibly the cause \citep{slk2016}. It is one of four pulsars in NG9WN where $\sigma_\const$ was less than the scintillation-estimated $\sigma_\DISS$ (at 820~MHz for this pulsar). Therefore, we are unable to determine the breakdown of $\sigma_\const$ into $\sigma_\J$ and $\sigma_\DISS$ in the band. We then assumed that the entirety of $\sigma_\const$ is comprised of $\sigma_\J$. NG9WN found that for most pulsars $\sigma_\J > \sigma_{\DISS}$ except at lower frequencies (e.g.~430~MHz) and so we believe our assumption to be reasonably accurate, though with a moderately high DM of $62.4~\mathrm{pc~cm^{-3}}$, DISS may play a larger role in the WN (as true for PSR~B1937+21, with $\mathrm{DM} = 71.0~\mathrm{pc~cm^{-3}}$). Thus our estimates of AEN and CEN, of which a large fraction is red, will be biased high.

\subsection{PSR~J1713+0747}

PSR~J1713+0747 is considered one of the best-timed pulsars, though red noise has been measured in the pulsar previously \citep{sns+2005,Zhu+2015,Lentati+2016}. The red noise was measured in \citet{Zhu+2015} similarly to the method in NG9, where the power-law noise parameters were $\Ared = 0.025~\mathrm{\mu s~yr^{1/2}}$ and $\gammared = -2.92$. Using Eq.~\ref{eq:sigmaTN2} with a timespan of 8.8~years equal to the observation length in the NG9, and integrating the implied power spectrum with the appropriate transmission function $\mathcal{T}(f)$, we calculated $\sigma_{\red} = 120$~ns. The two-dimensional likelihood surface of $\Ared$ and $\gammared$ in \citet{Zhu+2015} suggests that there can be large errors on both the amplitude and the spectral index, leading to a large fractional error on the rms red noise. Our analysis concludes that $\sigma_A = 119_{-5}^{+4}$~ns, which is consistent with the previous measurement, however we find no evidence of redness from the SF analysis.

Since PSR~J1713+0747 was observed with both AO and GBT, we performed our variance analysis on the residuals from each telescope separately after we accounted for the full timing model (i.e., both telescopes used in the fit). The results are listed in Table~\ref{table:summary}. We measured 4699 TOAs with AO and 10530 with GBT, the significant increase corresponding to the GUPPI backend at GBT coming online before the PUPPI backend at AO. The GBT data show significantly higher excess noise as well as a detection of CEN; the chromatic noise average into an upper limit when jointly analyzing the data. The CEN could result from the fact that the observing frequency range at GBT is lower than that at AO (1400 and 820~MHz bands versus 1400 and 2300~MHz) and therefore chromatic noise contributes more at GBT. It is interesting to note that note only is AO more sensitive, the excess noise contribution is less than that of GBT, though the exact cause is currently unknown.

\subsection{PSR~J1903+0327}

PSR~J1903+0327 is the pulsar with the highest DM ($297.6~\mathrm{pc~cm^{-3}}$) in the NANOGrav pulsar timing array and shows the largest value of CEN that is strongly red, suggesting some influence from the ISM. It shows a peak in the log-variance likelihood for AEN although we report an upper limit based on our cutoff criterion. The large DM correlates with small scintillation bandwidths and timescales, leading to a large number of scintles across the observing band and a low probability of bright scintillation events. Therefore, estimating $\sigma_\J$ in the future may be difficult. Like PSR~J1643$-$1224, NG9WN also estimated $\sigma_\const$ as less than the scintillation-estimated $\sigma_\DISS$ for PSR~J1903+0327 (at 1400~MHz for this pulsar). Again, we assume that the entirety of $\sigma_\const$ is comprised of jitter for consistency though the assumption may be incorrect; the uncertainty will bias our red noise estimates. We report a CEN measurement of $\sigma_C = 3.2_{-0.3}^{+0.2}~\mu\mathrm{s}$ for this pulsar although it is observed with the 1400 and 2300~MHz receivers at AO. Scaled to the mid-range frequency of $\sim$1800~MHz observed for this pulsar and assuming $\s = 2$ we find $\sigma_{C,1800} = 1.0 \pm 0.1~\mu\mathrm{s}$.

%1.775 GHz: 1.02-0.10+0.06
%1.8 GHz: 0.99-0.09+0.06

\subsection{PSR~J1909$-$3744}

PSR~J1909$-$3744 is one of the best-timed pulsars known. It shows the lowest AEN upper limit for all pulsars and was observed for the entire $\sim$9~year timespan. It is the pulsar with the second-ranked total WN in NG9WN, behind PSR~B1937+21, which is further confirmed in Figure~\ref{fig:acwsummary}. Due to the lack of AEN, the pulsar is the most intrinsically stable clock in the NANOGrav pulsar timing array. If the CEN can be mitigated in the future, it may become the overall most stable clock in our pulsar timing array.

\subsection{PSR~J1918$-$0642}

As with several other pulsars, we see strong peaks in the log-variance likelihoods for both AEN and CEN (and a detection in the single-process excess). The $\sigma_\J$ measurement from NG9WN is an upper limit, which means that our method causes excess noise to be absorbed by the WN. The large value of $\niss$ measured in the 820~MHz band implies that even after $9$~years of timing, achieving a large pulse $\SN$ to measure pulse jitter may be difficult.

\subsection{PSR~B1937+21}
\label{sec:B1937+21}

Timing residuals for PSR~B1937+21 show the most obvious signature of red noise due to the fact that the AEN, the vast majority of which is red from the SF analysis, dominates over the WN. This pulsar is the only one in the pulsar timing array where the AEN dominates. Sources of the achromatic red noise have been attributed to a variety of causes, including: neutron star rotational irregularities (SC2010), spin perturbations from an orbiting asteroid belt \citep{sc+2013}, and GWs. The chromatic component is thought to be due to interstellar propagation effects \citep{Armstrong1984,Rickett1990,ktr1994,cbl+1995}, such as DM and scattering variations \citep{cwd+1990}. \citet{Lentati+2016} found evidence for both achromatic and chromatic noise components in the International Pulsar Timing Array Data Release 1. While the ISM has been suggested as a source of red noise, we find no evidence for any leakage of chromatic noise into the residuals for PSR~B1937+21 at the $\sim$70~ns level (95\% upper limit) over 9.1~years of timing. However, our timing model includes a large number of $\Delta \DM_i$ bins spanning that time and the most $\FD$ parameters (5) of any pulsar's model in our data set which can absorb CEN components.

As with PSR~J1713+0747, we analyzed the residuals from each telescope independently and find that the excess noise at AO (from 2513 analyzed TOAs) is also slightly lower than that at the GBT (from 7217 analyzed TOAs). From our SF analysis, we see that the majority of the excess noise for PSR~B1937+21 is red. It is unclear if the contributing red processes are common to the pulsar and/or the line of sight path, however the large majority of the variance should be from observatory-independent causes. Since both sets of residuals from each telescope derive from one global timing model fit, it is unclear how each telescope contributes to the timing model parameter estimation and therefore the post-fit excess noise as well.

\subsection{PSR~J2010$-$1323}

PSR~J2010$-$1323 shows a strong peak in the log-variance likelihoods of AEN though it does not meet our cutoff for detection, which is also likely biased from the upper limit on WN. We compute an upper limit in CEN. The pulsar weakly scintillates and therefore it will be difficult to constrain $\sigma_\J$, especially at the 820~MHz receiver band where the number of scintles observed is large.

\subsection{PSR~J2017+0603}

PSR~J2017+0603 was observed in three different receiver bands in NG9 and shows upper limits in both red noise components. The WN parameters at 430~MHz and 1400~MHz are not well-constrained though less than 2~years of data have been collected. The pulsar strongly scintillates at 1400~MHz and therefore a measurement of $\sigma_\J$ should be possible in the near future. At 430~MHz, the number of scintles is large, and coupled with the low pulse $\SN$, estimating jitter will be difficult. Data from the 430~MHz receiver were collected for only about one third of the observing timespan and is currently no longer observed using the receiver. Therefore, refining the noise model for the receiver band is not a priority and the data may be excluded from future timing analyses.

\subsection{PSR~J2317+1439}

PSR~J2317+1439 contains the only 327~MHz data in NG9. The pulsar was observed with both 327~MHz and 430~MHz for the majority of its total timespan until the last two years when 1400~MHz observations were added. We detect AEN in the timing residuals of the pulsar, and since the white noise parameters were well-determined for all three bands, we believe the measurement to be accurate.

\subsection{Low $\SN$: PSRs~J1747$-$4036, J1832$-$4036, J1949+3106}

We excluded these pulsars from our analyses because all pulses in at least one receiver band did not meet the $\SN < 3$ requirement for inclusion in the NG9WN analysis (note that the time and frequency resolutions were not the same as in this work). Therefore, the WN model is undetermined for the pulsars in those specific receiver bands and we are unable to adjust the errors on the TOAs accordingly.

\subsection{Instrumental Error: PSRs~J1741+1351, J1923+2515, B1953+29}

We removed timing residuals between MJDs 55750 and 56020 corresponding to an instrumental error with 430~MHz data taken by the ASP receiver for these three pulsars only. PSR~B1953+29 still shows chromatic excess noise though the timespan over which dual frequency data remain is only $\sim$1~year as compared to the total 7.2~years observed; the largely red noise may then be in part to poor DM estimation. Similarly, only $\sim$1~year of dual frequency data remain for PSR~J1741+1351, resulting however in an upper limit on CEN. PSR~J1923+2515 is the pulsar observed for the shortest timespan (1.6~years) that shows a detection of AEN.

\subsection{Single-Frequency Data with a Single $\DM_i$ Parameter: PSRs~J1910+1256, J1944+0907}

PSRs~J1910+1256 and J1944+0907 show detections of strongly red AEN but have long spans of early single-frequency-band (1400~MHz) coverage with a much smaller frequency range using the ASP backend (for PSR~J1944+0907, 430~MHz data are present for the first epoch only). A single DM was fit over a significant fraction of the total timespan: 1650 days for PSR~J1910+1256, 1250 days for PSR~J1944+0907. We therefore expect errors in the timing residuals from the time-varying DM \citep[e.g.,][]{Lam+2015}, a chromatic effect. However, because the ASP bandwidth is so small, our method cannot separate chromatic noise from achromatic noise. Some large portion of the measured AEN is then actually CEN from unmodeled DM variations. Without legacy data from those time periods, those data may need to be removed from the data set for robust GW detection.

\subsection{Additional Pulsars with Disagreement Between $\sigma_{\const}$ and $\sigma_{\DISS}$: PSRs~J1600$-$3053, J1640+2224} 

Like PSRs~J1643$-$1224 and J1903+0327, these two pulsars (PSR~J1640+2224 at 430~MHz and PSR~J1600$-$3053 at 820~MHz) showed a disagreement between the upper limits on $\sigma_{\const}$ and the scintillation-estimated $\sigma_{\DISS}$. We repeat the same procedure for adjusting the WN parameters and both show significant chromatic red noise.

\subsection{AEN Detections, CEN Upper Limits: PSRs~J0030+0451, J0613$-$0200, J1744$-$1134, B1855+09, J2145$-$0750, J2214+3000}

These pulsars show detections of AEN and upper limits of CEN, four of which show strongly red residual: PSRs~J0030+0451, J0613$-$0200, B1855+09, and J2145$-$0750), with J1744$-$1134 showing a significant detection of red noise. PSR~J2214+3000 was only observed for $\sim$2~years while the rest were observed for $\sim$9~years. PSR~J0613$-$0200 shows an AEN detection despite an upper limit in the WN model for both receiver bands. The large number of scintles $\niss$ implies that lack of strong scintillation events may make refinement of the WN parameters, specifically $\sigma_\J$, difficult in the future given the large quantity of data already available for the pulsar. The AEN value is therefore biased for PSR~J0613$-$0200. \citet{Reardon+2016} find steep red noise in this pulsar $P(f) \sim f^{-5}$ and our large value of RP ($9.3 \pm 2.4$, the third highest value) shows some loose agreement. PSR~J0030+0451 has the third-highest value of $\RP = 9.0 \pm 2.5$, which also loosely suggests some agreement with NG9 ($P(f) \sim f^{-6}$)

\subsection{Noise Parameter Upper Limits: PSRs~J0931$-$1902, J1614$-$2230, J1738+0333, J1853+1303, J2043+1711, J2302+4442}

These pulsars show upper limits in both the excess noise and the WN parameters. In the case of PSRs~J0931$-$1902, J2043+1711, and J2302+4442, the observed timespans are $\lesssim 2$~years, with PSR~J0931$-$1902 on the short end (0.6~years). It is therefore unsurprising that the noise parameters are ill-constrained for these pulsars. 

The other three pulsars were observed for between 4 and 6 years but show no hints of excess noise. The WN parameters for the pulsars are not well-constrained in at least one of the receiver bands and therefore the conservative errors may be absorbing some of the excess noise. Even so, it is possible that some subset of the pulsars may turn out to be stable clocks on longer timescales. A large value of $\niss$ was observed for PSR~J1614$-$2230 in NG9WN, and therefore it may be difficult to estimate $\sigma_\J$ for the 820~MHz band. On the reverse side, PSR~J1738+0333 strongly scintillates and therefore $\sigma_\J$ may be estimated at 2300~MHz in the near future. PSR~J1853+1303 shows strong scintillation at 1400~MHz but weaker potential for scintillation events at 430~MHz. The lower receiver band was only observed over the second half of the data span however so it is possible that increased observing will help improve both sets of WN parameters.

\begin{deluxetable*}{ccccccc}
\tablewidth{0pt}
\tablecolumns{7}
\tablecaption{Best-fit Excess Noise Parameters}
\tablehead{
\colhead{Fit} & \colhead{$\ln C_2$} & \colhead{$\alpha$} & \colhead{$\beta$} & \colhead{$\gamma$} & \colhead{$\delta$} & \colhead{$N_{\rm D}(N_{\UL})^{\rm a}$}
}
\startdata
CP & $2.1 \pm 0.3$ & $-0.85 \pm 0.09$ & $0.99 \pm 0.03$ & $1.86 \pm 0.09$ & $1.54 \pm 0.05$ & 575(458)\\
MAG & $0.4_{-5.6}^{+5.4}$ & $-2.7_{-2.5}^{+2.4}$ & $1.4 \pm 0.4$ & $2.4_{-0.6}^{+0.7}$ & $1.9_{-0.4}^{+0.5}$ & 15(7)\\
MSP$_{10,\mathrm{PPTA}}$ & $0.7 \pm 2.7$ & $-1.2 \pm 0.5$ & $0.8 \pm 0.1$ & $2.2 \pm 0.3$ & $1.0 \pm 0.1$ & 27(147)\\
NANO & $-16_{-17}^{+14}$ & $0.8_{-2.0}^{+2.2}$ & $0.9_{-0.6}^{+0.7}$ & $4.9_{-2.6}^{+4.9}$ & $1.7_{-0.7}^{+1.2}$ & 9(25)\\
CP+MAG & $2.5 \pm 0.3$ & $-1.35 \pm 0.08$ & $1.11 \pm 0.04$ & $1.67 \pm 0.08$ & $1.64 \pm 0.05$ & 590(465)\\
CP+MSP$_{10,\mathrm{PPTA}}$ & $1.8 \pm 0.3$ & $-1.31 \pm 0.04$ & $1.08 \pm 0.03$ & $1.95 \pm 0.08$ & $1.56 \pm 0.05$ & 602(605)\\
MSP$_{10,\mathrm{PPTA}}$+NANO & $-1.2 \pm 2.5$ & $-0.9 \pm 0.4$ & $0.8 \pm 0.1$ & $2.3_{-0.2}^{+0.3}$ & $1.1 \pm 0.1$ & 36(172)\\
CP+MSP$_{10,\mathrm{PPTA}}$+NANO & $1.8_{-0.3}^{+0.2}$ & $-1.32 \pm 0.04$ & $1.08 \pm 0.03$ & $1.96 \pm 0.08$ & $1.56 \pm 0.05$ & 611(630)\\
CP+MAG+MSP$_{10,\mathrm{PPTA}}$ & $2.5 \pm 0.2$ & $-1.43 \pm 0.04$ & $1.11 \pm 0.03$ & $1.73 \pm 0.08$ & $1.62 \pm 0.05$ & 617(612)\\
CP+MAG+MSP$_{10,\mathrm{PPTA}}$+NANO & $2.5 \pm 0.2$ & $-1.42 \pm 0.04$ & $1.10 \pm 0.03$ & $1.73 \pm 0.08$ & $1.62 \pm 0.05$ & 626(637)
\enddata
\label{table}
\tablenotetext{}{The model used for discussion in the text and subsequent figures is the final one, which includes all pulsars in the fit.}
\tablenotetext{a}{$N_{\rm D}$ are the number of detected measurements of excess noise while $N_{\UL}$ are the number of upper limit measurements.}
\end{deluxetable*}

%\begin{landscape}
\begin{figure*}[t!]
\epsscale{1.2}
\plotone{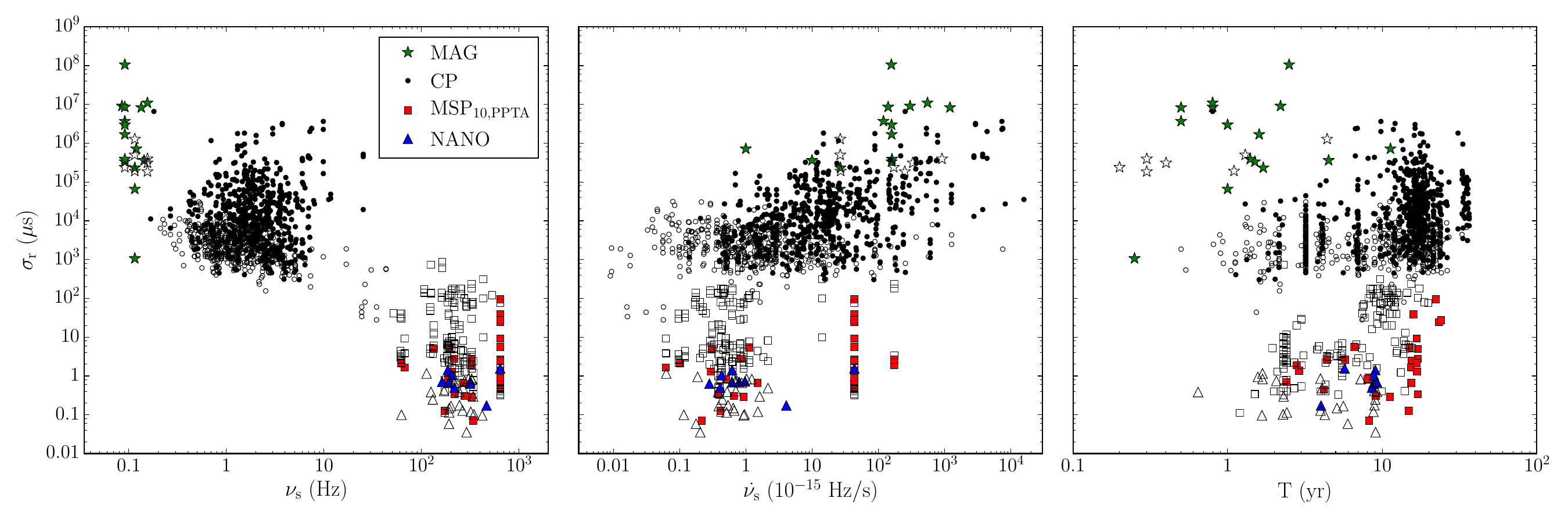}
  \caption{\footnotesize Measured red noise versus spin frequency (left), spin frequency derivative (middle), and observing timespan (right). Green stars indicate magnetars (MAGs), black circles indicate canonical pulsars (CPs), and red squares indicate millisecond pulsars (MSPs), all from \citet{sc2010}. Blue triangles indicate NANOGrav MSPs (NANO) with measurements derived from this work. Filled symbols indicate detections of red noise while unfilled symbols indicate upper limits.}
\label{fig:fig2_comparison}
\end{figure*}
%\end{landscape}

\section{Global Scaling Law Maximum Likelihood Analysis}
\label{sec:MLA}

Using our AEN measurements, we re-examined the spin-noise scaling relations of SC2010 across different pulsar populations: magnetars (denoted MAG), slow-period canonical pulsars (denoted CP), and MSPs. SC2010 had measured spin noise for only two MSPs (though multiple measurements per pulsar were allowed), making the NANOGrav pulsars important additions for anchoring the global fit. We estimated the amount of achromatic red noise by adjusting our AEN measurements using the RP values from \S\ref{sec:SF}. While the RP values characterize the total variance and not purely the achromatic noise, the process of epoch-averaging the residuals for the SF analysis reduces the chromatic noise as roughly the square root of the number of observing bands (typically two). Nevertheless, the bias is still present and introduces scatter into our measurements. We defined a red noise ``detection'' as when the AEN was detected and the estimate of $V_\red/V_{\rm tot}$ was ``$2\sigma$'' significant. Otherwise, a detection of AEN was treated as an upper limit on the achromatic red noise. For 95\% upper limits on AEN, we treated those as upper limits on the red noise as well.

We also incorporated red noise results from 15 out of 20 MSPs in the Parkes Pulsar Timing Array (PPTA) extended first data release \citep{Manchester+2013}. \citet{Reardon+2016} modeled the achromatic red noise with a modified power law of the form
\be
P(f) = \frac{P_0}{\left[1+\left(f/f_c\right)^2\right]^{\alphared/2}},
\ee
where the amplitude of the power $P_0$ at a corner frequency of $f_c$ is in units of yr$^3$ and $\alphared$ is the spectral index. We calculated the rms red noise for each pulsar by integrating its power spectrum after applying the appropriate transmission function $\mathcal{T}(f)$ (see Eq~\ref{eq:postfit}). 

For our analysis, we assumed that the entirety of the achromatic red noise is due to intrinsic spin noise. Other sources of achromatic red noise, such as GWs, will bias our measurements. Excess WN from improper adjustment of the AEN will also bias our measurements. Any other pulsars for which red noise was measured in SC2010 that contain chromatic noise will also skew the results.

As in SC2010, we assumed that the red spin noise is a function of $\nu_{\rm s}$ and $\dot{\nu}_{\rm s}$, the pulsar spin frequency and spin frequency derivative, respectively, as well as a function of observing timespan $T$. For all MSPs, we applied Shklovskii corrections to $\dot{\nu}_{\rm s}$ where possible \citep{Shklovskii1970,nt1995}. Since the red noise $\sigma_\red$ was attributed to rotational spin noise alone, $\sigma_\red$  is entirely a function of the two basic pulsar spin observables and time. SC2010 argues that for most pulsars, spin noise is the dominant source of excess noise. Again, at the timing precision level of MSPs, it is not clear that spin noise should be the dominant term. However, while the MSP spin noise values will be biased, the global trends in overall red noise remain the same. Therefore, we believe that using the three-observable formalism with parameters that only represent spin noise is adequate to describe the red noise in general though more parameters may be required in the future. For example, if a trend is present in Figure~\ref{fig:red_vs_DM}, then a scaling in DM should be included in the fit.

Figure~\ref{fig:fig2_comparison} shows the red noise as a function of the three observables. The different pulsar populations are displayed with differently shaped symbols and we explicitly separate the NANOGrav pulsars from the other MSPs provided in SC2010 and \citet{Reardon+2016}, denoted as MSP$_{10,\mathrm{PPTA}}$. Detections of red noise are given by the filled symbols and upper limits are given by the unfilled ones.

Following SC2010, we write the rms red noise after a timing-model fit as
\be
\hat{\sigma}_\red(\nu_{\rm s},\dot{\nu_{\rm s}},T|C_2,\alpha,\beta,\gamma) = C_2 \nu_{\rm s,Hz}^\alpha |\dot{\nu}_{\rm s,-15}|^\beta T_{\rm yr}^\gamma~\mathrm{\mu s}
\label{eq:hatsigex}
\ee
where $C_2, \alpha, \beta$, and $\gamma$ are parameters to be estimated over each pulsar included in the analysis. We include a parameter $\delta$ that accounts for the scatter in the red noise amplitude. For detections of red noise, we assume that the variance is log-normally distributed, represented with a PDF
\ba
\!\!\!\!\!\!& f_{\rm D} (\sigma_\red|\{\nu_{\rm s},\dot{\nu}_{\rm s},T\},\{C_2,\alpha,\beta,\gamma,\delta\}) = \quad\quad\quad\quad\quad\quad\quad \nonumber \\
\!\!\!\!\!\!& \displaystyle{\frac{1}{\sqrt{2\pi\delta^2}}}  \exp\!\left(\!-\frac{\displaystyle{[\ln\sigma_\red - \ln\hat{\sigma}_\red(\nu_{\rm s},\dot{\nu}_{\rm s},T|C_2,\alpha,\beta,\gamma)]}^2}{2\delta^2}\right)\!\!.
\ea
For upper limit measurements on the excess noise, we use the survival function of the log-normal distribution (e.g., see \citealt{Lawless} for a review, noting that an upper limit likelihood is equivalent to a left-censored distribution, or 1 minus the survival function) to calculate the appropriate distribution for upper limits as
\ba
& f_{\UL}(\sigma_\red|\{\nu_{\rm s},\dot{\nu}_{\rm s},T\},\{C_2,\alpha,\beta,\gamma,\delta\}) = \quad\quad\quad\quad\quad\quad\quad \nonumber \\
& \displaystyle{\frac{1}{2} \left(1+\mathrm{erf}\left[\frac{\ln\sigma_\red - \ln\hat{\sigma}_{\red}(\nu_{\rm s},\dot{\nu}_{\rm s},T|C_2,\alpha,\beta,\gamma)}{\delta \sqrt{2}} \right]\right)},
\ea
where erf is the error function. We note that the order of the argument to erf presented here corrects a typo in SC2010, which is reversed by a negative sign.

We define our five-parameter likelihood function as
\ba
&& \Like(C_2,\alpha,\beta,\gamma,\delta | \{\sigma_{\red,i},\nu_{\mathrm{s},i},\dot{\nu}_{\mathrm{s},i},T_i\}) \nonumber \\
&&=\prod_j f_{\rm D}(\sigma_{\red,j}|\{\nu_{\mathrm{s},j},\dot{\nu}_{\mathrm{s},j},T_j\},\{C_2,\alpha,\beta,\gamma,\delta\}) \nonumber \\
&& \times\prod_k f_{\UL}(\sigma_{\red,k}|\{\nu_{\mathrm{s},k},\dot{\nu}_{\mathrm{s},k},T_k\},\{C_2,\alpha,\beta,\gamma,\delta\})
\label{eq:param_likelihood}
\ea
where $i$ labels all of the individual measurements, $j$ labels the subset of measurements with detected red noise, and $k$ labels the subset of measurements with upper limits on the red noise. We performed a linear grid search over the five parameters $\ln C_2, \alpha, \beta, \gamma, \delta$ with range and resolution to sufficiently sample the parameter space. For computational stability, we first computed the log-likelihood as the sum of the log-PDFs in Eq.~\ref{eq:param_likelihood}.

\begin{figure}[t!]
\hspace{-5ex}
\includegraphics[width=0.55\textwidth]{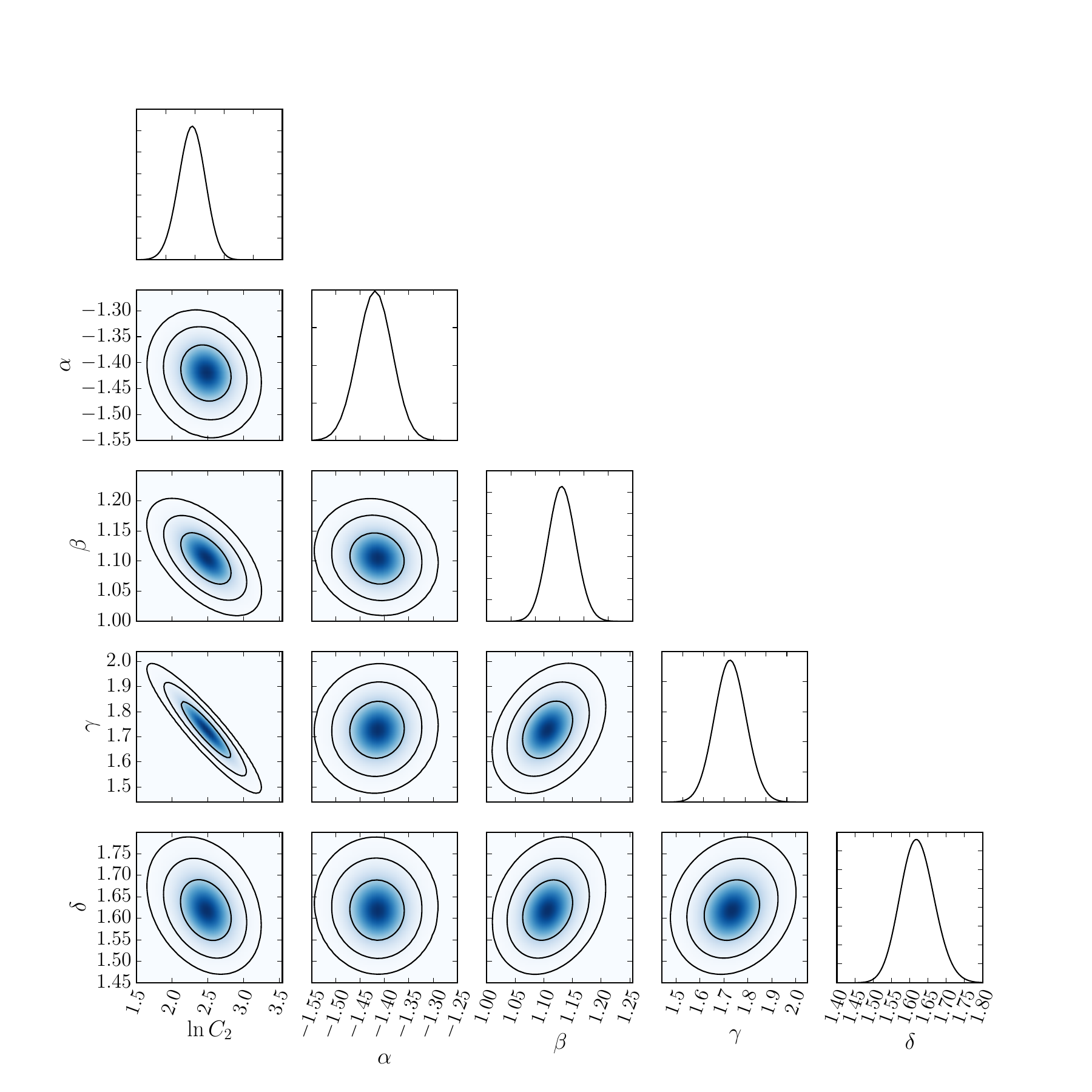}
  \caption{\footnotesize Global spin noise maximum likelihood analysis results when computed for all pulsars. The colored regions show the two-dimensional marginalized distributions, with the contours encapsulating the 68.3\%, 95.5\%, and 99.7\% of the total probability. The line plots show the one-dimensional marginalized distributions from which the parameter estimates are calculated.}
\label{fig:likelihood_ALL}
\end{figure}

Figure~\ref{fig:likelihood_ALL} shows the likelihood analysis run with all pulsars included. Parameters were estimated by taking the one-dimensional marginalized distributions, computing the cumulative distribution function, and finding the $\pm$34.1\% confidence regions about the maximum likelihood value. We performed a similar analysis over different subsets of the pulsar populations and provide our results in Table~\ref{table}. 

\begin{figure}[t!]
\hspace{-5ex}
\includegraphics[width=0.5\textwidth]{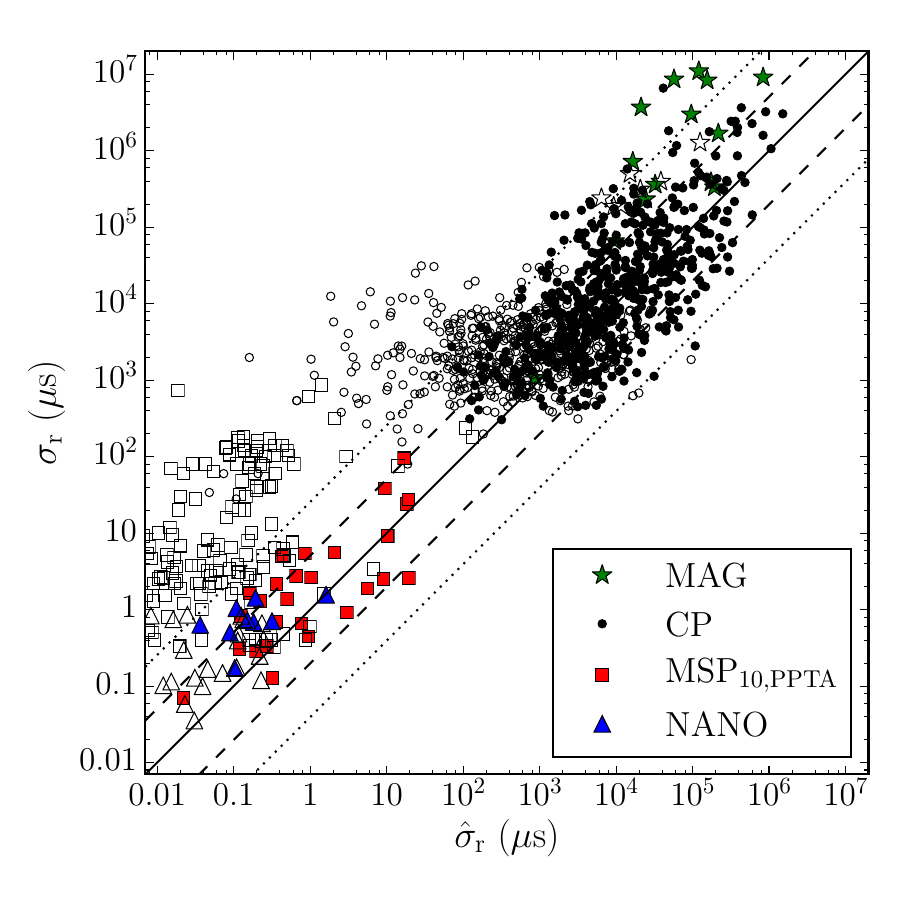}
  \caption[Measured excess noise versus model excess noise]{Measured excess noise versus model excess noise from Eq.~\ref{eq:hatsigex} and using global parameters fit for over all pulsars as given in Table~\ref{table}. Symbols are as defined in Figure~\ref{fig:fig2_comparison}, with filled symbols being detections and open symbols being upper limits. The solid line shows where the measured excess noise equals the model excess noise, and the dashed and dotted lines show the 1 and $2\sigma$ errors as estimated by $\delta$, respectively.}
\label{fig:fig4_comparison}
\end{figure}

Figure~\ref{fig:fig4_comparison} shows the measured red noise versus the model red noise using the maximum likelihood model parameters when computed for all pulsars. The addition of the NANOGrav and PPTA pulsars to the fit helps constrain the power law relations at lower $\nu_{\rm s}$ and $\dot{\nu}_{\rm s}$ as compared with SC2010, which only contained (multiple estimates of) detected noise in two MSPs. We note that the global fit tends to underestimate the spin noise in the NANOGrav pulsars slightly as compared with the total MSP population; the cause could be from the differences in timing models between NANOGrav and other measurements but it is unclear from the values alone. In addition, the values for magnetars are also underestimated, suggesting either that systematic bias is present when trying to fit over all pulsar populations or that the different pulsar populations are not fit by a single scaling law.

For the NANOGrav-only data (NANO), we find that our three scaling parameters \{$\alpha$, $\beta$, $\gamma$\} are all roughly consistent with zero, reflecting that the red noise measurements by themselves are generally unconstraining. In terms of the scaling with time $\gamma$, we also suffer from the fact that most of our detections of red noise are in pulsars that have been observed for $\sim 9$ years, making constraints on $\gamma$ difficult. For the combined MSP$_{10,\mathrm{PPTA}}$ and NANO model, we see steeper dependence with time than the global fit. The largest values of $\sigma_{\red}$ (detections $\gtrsim 10~\mu\mathrm{s}$), with the largest $T$, are entirely from PSR~B1937+21, suggesting that the pulsar alone is driving the steep fit. How representative the pulsar is of the MSP population is currently unclear (see \S\ref{sec:B1937+21}).

\section{Conclusions}
\label{sec:conclusions}

We examined the amount of post-timing-model-fit excess noise in NANOGrav timing residuals beyond the measured WN. After re-weighting our TOAs, we detected excess noise in 26 of our pulsars. Splitting the excess noise into achromatic and chromatic components as well as determining the redness of the power spectra of the residuals for each pulsar will help us in attributing the proper amount of noise to different physical sources. Using the achromatic red noise estimates, we were then able to further bound the global pulsar scaling relations for spin noise. While chromatic noise is possibly correctable, as well as any achromatic white noise, it is unclear if that is true for the achromatic red noise, and by how much.

The addition of measurements from the NANOGrav pulsars helps constrain the spin noise scaling relations first derived in SC2010 by anchoring the fit at lower $\nu_{\rm s}$ and $\dot{\nu}_{\rm s}$. The global scaling relation derived from using all of the pulsars will be minimally affected by uncertainties in our red noise estimates due to the many orders of magnitude spanned in the parameters for all of the pulsars. For a process with power spectrum $P(f) \propto f^{-\gammared}$, the spectral index $\gammared$ is related to the spin noise scaling parameter $\gamma$ by $\gammared = 2\gamma+1$ \citep{NG9yr}. Therefore, for $\gamma = 1.73 \pm 0.08$ from the fit over all pulsar populations, noise in residuals described with a power-law spectrum will have spectral index $\gamma_{\rm red} = 4.46 \pm 0.16$. It is not clear from our analysis if such a scaling law applies to individual pulsar populations, and while the MSP-only fit (both MSP$_{10,\mathrm{PPTA}}$ and NANO) suggests even larger values of spin noise in time, it is possible that PSR~B1937+21 alone is dominating our fit, potentially biasing the population parameter estimate. However, we expect GWs to contribute to our red noise estimates and so it is likely that some component of our achromatic noise is due to GWs; a power-law GW background will cause timing residuals with a $\gammared = 13/3$ spectrum although recent work finds evidence for a deficit of power at low fluctuation frequencies \citep{NG9GWB}. Understanding how the different sources of red noise in MSPs scale with time is crucial for time-to-detection estimates for GWs and for understanding long-term usability of MSPs in a pulsar timing array. Improved estimation of the red noise in the future, either with new measurements or with refinements of the current measurements (e.g., the decomposition into red and white), is of vital importance.

As demonstrated by the Laser Interferometer Gravitational-Wave Observatory's recent detection of GWs, understanding the noise within the detector is imperative for the instrument's capability to detect GWs \citep{LIGO2016,LIGONoise2016}. With NANOGrav and other collaborations moving towards the detection of low-frequency GWs, we require full noise characterization of our pulsar timing arrays. Implementation of the full noise model we describe will help remove biases on timing parameter estimation and improve sensitivity towards the growing GW signal in our data.

\acknowledgments

{\it Author contributions.} MTL developed the mathematical framework, created the modified data set and residuals, undertook the analysis, and prepared the majority of the text, figures, and tables. JMC and SC helped with the development of the framework, the format of figures and tables, and generated some additional text. RMS provided useful guidance with the data and interpretation of results in SC2010. JAE developed the noise model in NG9 and assisted in comparison with the previous noise analysis results, as well as provided a forum for discussion of NANOGrav's noise budget along with DRM and XS. ZA, KC, PBD, TD, JAE, RDF, EF, MEG, GJ, MJ, MTL, LL, MAM, DJN, TTP, SMR, IHS, KS, JKS, and WWZ all ran observations and developed timing models for the NG9 data set. Additional specific contributions are described in \citet{NG9yr}.

{\it Acknowledgments.} We thank Paul Baker for useful discussion regarding the methods. The NANOGrav Project receives support from NSF PIRE program award number 0968296 and NSF Physics Frontier Center award number 1430284. NANOGrav research at UBC is supported by an NSERC Discovery Grant and Discovery Accelerator Supplement and the Canadian Institute for Advanced Research. MTL acknowledges partial support by NASA New York Space Grant award number NNX15AK07H. JAE acknowledges support by NASA through Einstein Fellowship grant PF3-140116. Portions of this research were carried out at the Jet Propulsion Laboratory, California Institute of Technology, under a contract with the National Aeronautics and Space Administration. TTP was a student at the National Radio Astronomy Observatory (NRAO) while this project was undertaken. Data for the project were collected using the facilities of the NRAO and the Arecibo Observatory. The NRAO is a facility of the NSF operated under cooperative agreement by Associated Universities, Inc. The Arecibo Observatory is operated by SRI International under a cooperative agreement with the NSF (AST-1100968), and in alliance with the Ana G. M\'{e}ndez-Universidad Metropolitana, and the Universities Space Research Association.

\end{document}